\documentclass{article}
\usepackage[letterpaper,top=2cm,bottom=2cm,left=2cm,right=2cm,marginparwidth=1.75cm]{geometry}

\usepackage{mathtools}
\usepackage{subfig}
\usepackage{graphicx}
\usepackage{amssymb}
\usepackage{hyperref}
\newtheorem{example}{Example}

\title{Survey of Methods for Solving Systems of Nonlinear Equations, Part I: Root-finding Approaches}
\author{Ilias S. Kotsireas \\
ikotsire@wlu.ca \\
Wilfrid Laurier University  \\
Canada
\and
Panos M. Pardalos  \\
ppardalos@toxeus.org  \\
Toxeus Systems LLC \\
Orlando, Florida, USA
\and
Alexander Semenov \\
asemenov@ufl.edu \\
University of Florida \\
Gainesville, Florida, USA
\and
William T. Trevena \\
wtrevena@ufl.edu \\
University of Florida \\
Gainesville, Florida, USA
\and
Michael N. Vrahatis \\
vrahatis@math.upatras.gr \\
University of Patras \\
Patras, Greece
}
\begin{document}

\maketitle
\begin{abstract}
  This paper presents a comprehensive survey of methods which can be utilized to search for solutions to systems of nonlinear equations (SNEs). Our objectives with this survey are to synthesize pertinent literature in this field by presenting a thorough description and analysis of the known methods capable of finding one or many solutions to SNEs, and to assist interested readers seeking to identify solution techniques which are well suited for solving the various classes of SNEs which one may encounter in real world applications.
  
  To accomplish these objectives, we present a multi-part survey. In part one, we focus on root-finding approaches which can be used to search for solutions to a SNE without transforming it into an optimization problem. In part two, we will introduce the various transformations which have been utilized to transform a SNE into an optimization problem, and we discuss optimization algorithms which can then be used to search for solutions. In part three, we will present a robust quantitative comparative analysis of methods capable of searching for solutions to SNEs.
  
  

\end{abstract}




\maketitle


\section{Introduction}
This paper presents part one of a survey on methods for finding one or many solutions to a {\em system of nonlinear equations}\/ (SNE):
\begin{equation}\label{eq:1}
{F}_m(x) = {\mathit \Theta}_{m} \equiv (0, 0, \ldots, 0)^{\top} \quad \Longleftrightarrow \quad \left\{\begin{aligned}
    &f_1(x_1,x_2,\ldots,x_n)=0, \\[0.1cm]
    &f_2(x_1,x_2,\ldots,x_n)=0, \\
    &\kern1.45cm \vdots \\
    &f_m(x_1,x_2,\ldots,x_n)=0,
\end{aligned} \right.
\end{equation}
where ${F}_m=(f_1,f_2,\ldots, f_m): {\mathcal D}_n \subset {\mathbb R}^n
\to {\mathbb R}^m$,
where $f_1,f_2,\ldots,f_m$ are real-valued continuous or continuously differentiable functions on the domain ${\mathcal D}_n$, and where at least one of $f_1,f_2,\ldots,f_m$ is nonlinear. For example, consider the system of transcendental equations
\begin{equation}\label{eq:00}
    {F}_2(x) = {\mathit \Theta}_{2} \equiv (0, 0)^{\top} \quad \Longleftrightarrow \quad \left\{\begin{aligned}
    f_1(x_1,x_2)&=x_1-x_1\sin(x_1+5x_2)-x_2\cos(5x_1-x_2)=0, \\[0.1cm]
    f_2(x_1,x_2)&=x_2-x_2\sin(5x_1-3x_2)+x_1\cos (3x_1+5x_2)=0,
\end{aligned}  \right.
\end{equation}
which is comprised of two transcendental equations of two unknowns (See Figure 1). 

\begin{figure}[h]\label{fig1}
  \centering
  \includegraphics[width=\linewidth]{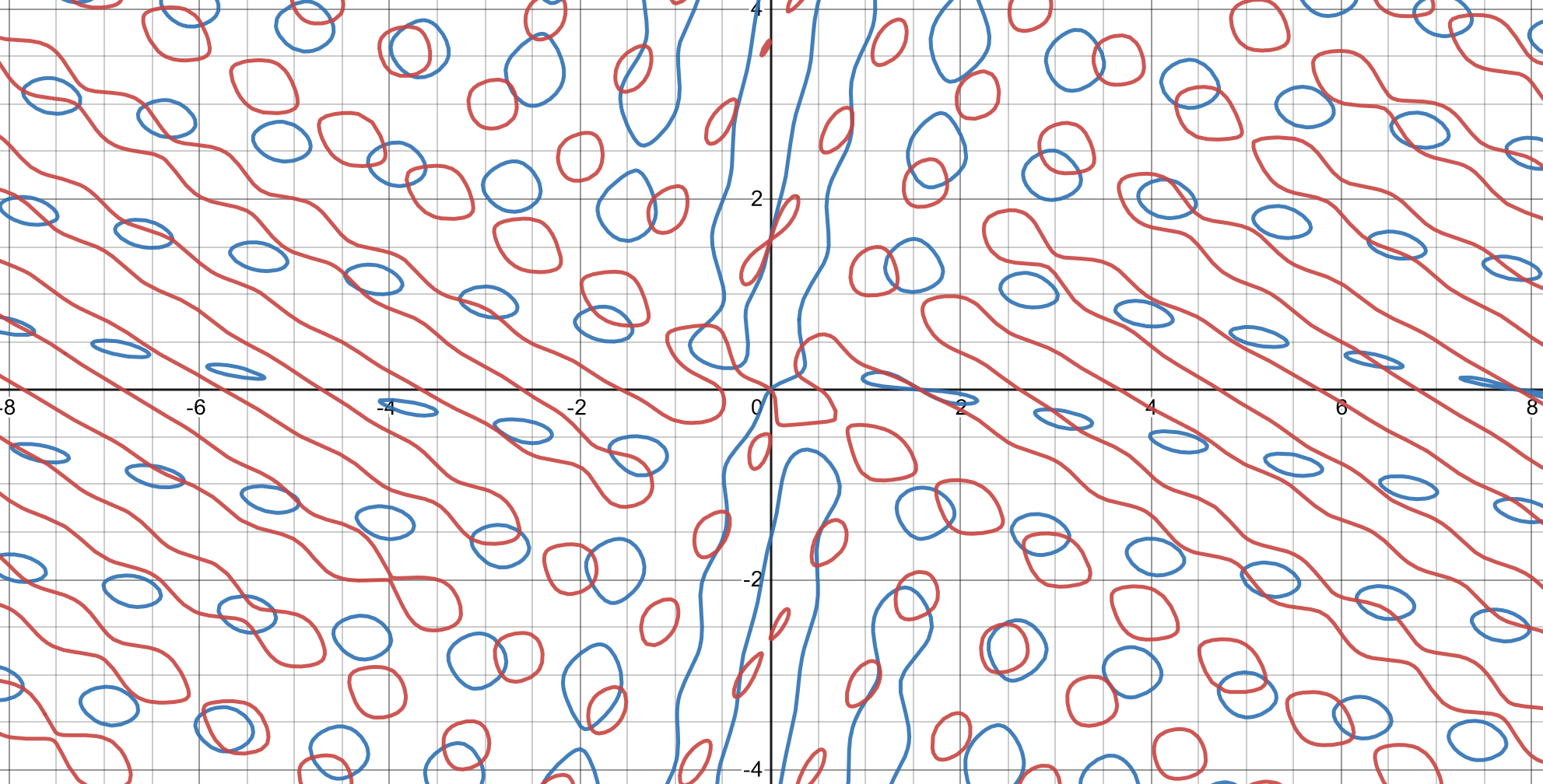}
  \caption{An example of a SNE with two transcendental equations of two unknowns as introduced by Eq.~(\ref{eq:00}): \\(Blue): $f_1(x_1,x_2) = x_1-x_1\sin(x_1+5x_2)-x_2\cos(5x_1-x_2)=0$; \ \ (Red): $f_2(x_1,x_2) = x_2-x_2\sin(5x_1-3x_2)+x_1\cos (3x_1+5x_2)=0$. Solutions to this SNE are defined as the points where the blue and red contours intersect. Finding all of the points within a certain region which satisfy both equations is a challenging task. 
  }
\end{figure}

Finding one or more solutions to a SNE is a challenging and ubiquitous task faced in many fields including chemistry \cite{FLOUDAS2000125,SACCO20115424,Holstad1999}, chemical engineering \cite{Jimenez-Islas2013},
automotive steering \cite{HENDERSON2010551},
power flow \cite{Chiang2014, mehta2016numerical}, large-scale integrated circuit designs \cite{Chiang2018},
climate modeling \cite{yang2010fully},
materials engineering \cite{Schneider2019}, robotics \cite{ZHANG2006,Lafmejani20151,cox2015robotics,Ji2003KinematicsAO}, 
nuclear engineering \cite{cht2016modification}, 
image restoration \cite{aji2020modified}, 
protein interaction networks \cite{Chiang2018}, neurophysiology \cite{verschelde1994}, economics \cite{Grosan2008}, 
finance \cite{golbabai2012}, applied mathematics \cite{zhang2009existence}, physics \cite{PhysRevLett.81.1195}, finding string vacua \cite{Mehta2011}, machine learning \cite{song2020nonlinear,pmlr-v108-cai20b}, geometric constraint solving (used in computer aided design) \cite{Gatilov2011PropertiesON}, and geodesy \cite{palancz2008dixon,Palancz2009} among others. 
The problem of solving even a system of polynomial equations has been proven to be NP-hard \cite{Jansson1998}. Furthermore, it has also been proven \cite{Matiyasevich} that no general algorithm exists for determining whether an integer solution exists for a polynomial equation with a finite number of unknowns and only integer coefficients. The latter has been known as Hilbert's 10th problem. 

\subsection{Notation / Scientific Style}
Throughout this paper, we utilize $x=(x_1, x_2, \ldots, x_n)^{\top} \in \mathcal{D}\subset \mathbb{R}^n$ to denote a real vector within the bounded domain $\mathcal{D}$. Furthermore, we utilize $x^*=(x_1^*, x_2^*, \ldots, x_n^*)^{\top} \in \mathcal{D}\subset \mathbb{R}^n$ to denote a real solution to a SNE such that all equations in the SNE are satisfied ($F_m(x^*)=0$). In an iterative method, we utilize $x^k=(x_1^k, x_2^k, \ldots,x_i^k,\ldots x_n^k)^{\top} \in \mathcal{D}\subset \mathbb{R}^n$ for $k=0,1,\ldots$ to denote the vector found during the $k-$th iteration of the iterative method. Here, $x^k_i$ denotes the $i-$th coordinate of the vector $x^k$.

\subsection{Terminology}
Although we refer to Eq.~(\ref{eq:1}) as a \textit{system of nonlinear equations} (SNE), such systems have been referred to in a variety of different ways in literature. For example, articles \cite{khirallah2013solving,khirallah2012novel,koupaei2015,el-shorbagy2020,naidu2016solving,kuri2003} utilize the abbreviation ``SNLE" to refer to a \textit{system of nonlinear equations}, and article \cite{xiao2018solving} uses the abbreviation ``SoNE". 
Other papers refer to Eq.~(\ref{eq:1}) as a \textit{nonlinear system of equations}, and use the abbreviations ``NSE" \cite{Geng2009,pei2019} and ``NLS"  \cite{Mousa2008}. Eq.~(\ref{eq:1}) has also been referred to as a \textit{nonlinear equation system} (NES) \cite{Gong2017,6849952,Qin2015,Guo2020b,LIAO2020113261,Liao2020,Gao2020,Gong2020c,Gao2021,HE2019104796,WU2021106733,Gao2021b,Song2020,LIAO2020105312,Cheng2018}. The survey in article \cite{gong2021nonlinear} uses the terminology \textit{Nonlinear Equations} (NEs) to refer to a system of one or more nonlinear equations.

When $m>n$, a SNE can be referred to as an {\em overdetermined}\/ SNE, and when $n>m$, a SNE can be referred to as an {\em underdetermined}\/ SNE. When $m=n$, a SNE can be referred to as a 
\textit{square}\/ SNE \cite{Ahookhosh2013}. Furthermore, a SNE is considered to be \textit{consistent}\/ if a solution exists which satisfies all equations~\cite{Gatilov2011PropertiesON}.

For a \textit{square}]/ SNE, a {\em solution $x^* =(x_1^*, x_2^*, \ldots, x_n^*)^{\top}$ of the}\/ SNE {\em ${F}_n(x) = {\mathit \Theta}_{n}$} or equivalently a {\em zero $x^*$ of the function ${F}_n(x)$}\/ or a {\em root $x^*$ of the function ${F}_n(x)$}\/ is called {\em simple}\/ if for the determinant of the corresponding {\em Jacobian matrix}:
\begin{equation}\label{jacobian}
 J_{F_{n}}(x) \equiv F'_n(x)_{ij} \equiv
\left\{\frac{\displaystyle\partial f_i(x)}{\displaystyle\partial x_j } \right\}_{ij} \equiv
\renewcommand{\tabcolsep}{2.5mm}
\renewcommand{\arraystretch}{1.6}
\left[ \begin{array}{cccc}
 \vspace*{0.1cm}
{\frac{\displaystyle\partial f_1(x)}{\displaystyle\partial x_1}} &
{\frac{\displaystyle\partial f_1(x)}{\displaystyle\partial x_2}} &
\cdots &
{\frac{\displaystyle\partial f_1(x)}{\displaystyle\partial x_{n}}}\\[0.2cm]
{\frac{\displaystyle\partial f_2(x)}{\displaystyle\partial x_1}} &
{\frac{\displaystyle\partial f_2(x)}{\displaystyle\partial x_2}} &
\cdots &
{\frac{\displaystyle\partial f_2(x)}{\displaystyle\partial x_{n}}}\\
 \vdots & \vdots & \ddots & \vdots\\[0.1cm]
{\frac{\displaystyle\partial f_n(x)}{\displaystyle\partial x_1}} &
{\frac{\displaystyle\partial f_n(x)}{\displaystyle\partial x_2}} &
\cdots &
{\frac{\displaystyle\partial f_n(x)}{\displaystyle\partial x_{n}}}  
\end{array} \right],\vspace*{0.1cm}
\end{equation}%
at $x^*$ it holds that ${\rm det}\,J_{F_{n}}(x^*) \neq 0$, otherwise it is called {\em multiple}.
The problem of conservation
and decomposition of a multiple root into simple roots
in the case of
systems of homogeneous algebraic equations has been tackled in~\cite{TanabeV2006}.
This approach can be applied to high dimensional
CAD where it is sometimes required to compute
the  
intersection of several hypersurfaces
that are a perturbation of a set of original unperturbed hypersurfaces.
 
When $F_n$ satisfies the {\em monotonicity condition}:
\begin{equation}\label{monotonicity}
\bigl(F_n(x)-F_n(y)\bigr)^\top (x-y) \geqslant 0,\kern0.3cm \forall\, x,y\in \mathbb{R}^n,
\end{equation}
the corresponding SNE can be referred to as a \textit{system of monotone nonlinear equations}~\cite{Ullah2021}. Furthermore, $F_n$ is considered to be Lipschitz continuous if there exists $L>0$ such that
\begin{equation}\label{l-continuity}
\|F_n(x)-F_n(y)\|_2\leqslant L\|x-y\|_2,\kern0.2cm \forall\, x,y \in \mathbb{R}^n.
\end{equation}
Many of the root finding methods described in Section \ref{rootfindingmethods} are guaranteed to converge to a solution when applied to SNEs that satisfy both the monotonicity and Lipchitz continuity conditions (the hybrid spectral methods introduced in \cite{aji2021} for example). 

\subsection{Comparison to other surveys}
Other surveys discussing solution techniques for SNEs include \cite{Martinez1994} and \cite{gong2021nonlinear}. We have decided to conduct this comprehensive literature review because many new solution techniques for SNEs have been introduced since the publication of  \cite{Martinez1994} in 1994, and because the recent survey presented in \cite{gong2021nonlinear} focuses mainly on 
methods which first convert a SNE into an optimization problem, and then search for multiple solutions to the optimization problem using \textit{Intelligent Optimization Algorithms} (IOAs). The IOAs discussed in article \cite{gong2021nonlinear} are primarily metaheuristics for global optimization. Although the survey in article \cite{gong2021nonlinear} provides a very nice discussion of IOAs for solving SNEs reformulated as optimization problems, many of the IOAs they discuss are only introduced at a very high level, only eight IOAs were tested in their computational study, and the IOAs were evaluated on SNEs comprised of 20 equations or less. Also, article \cite{gong2021nonlinear} only briefly mentions methods which can be used to search for solutions to SNEs without transforming them into optimization problems.

We would like to present a broader survey which covers in detail the large set of methods which can be used to solve a SNE without transforming it into an optimization problem (i.e.\ homotopy and symbolic computation methods). These methods are our main focus in part one of this survey. In part two, we will expand upon article \cite{gong2021nonlinear} by introducing additional reformulation techniques and optimization algorithms which have been used to solve SNEs, and by discussing in much more detail many optimization algorithms which were only briefly introduced in article \cite{gong2021nonlinear}. This will allow us to appropriately set the stage for the comprehensive empirical study we will present in part three of this survey. Furthermore, we believe it is imperative to introduce the reader to a technique for determining the number of solutions to a SNE that exist within a bounded domain. Such techniques are of critical practical importance for those interested in finding all solutions to a SNE that exist within a domain of interest.







\subsection{Organization of this survey}
We begin this survey by introducing a method which can be used to determine the total number of solutions to a SNE that exist within a given bounded domain. By determining the number of solutions to a SNE which exist within a bounded domain of interest, in the event that no solutions exist, one can avoid spending time and resources searching for solutions all together. Alternatively, if one knows that a specific number of solutions exist to a SNE within a bounded domain of interest, one can continue to search for solutions until the desired number of solutions are found. 
Next, 
we introduce root-finding methods which have been utilized in literature to search for solutions to a SNE without transforming it into an optimization problem.
We conclude our paper by introducing additional methods which have been used to attempt to solve SNEs, and 
by highlighting promising areas for future research. 

\section{Determining the number of solutions to a SNE in a bounded domain}\label{numsolutions}

\noindent
The knowledge of all the solutions of a system of nonlinear equations and/or all the extrema of a function is of major importance in various fields.
The total number of the solutions of a system of nonlinear equations can be obtained 
by computing the topological degree.
Suppose that the function
$F_n=(f_1, f_2, \ldots,f_n)\colon {\mathcal D}_n \subset $ ${\mathbb R}^n$ $\to $ ${\mathbb R}^n$
is defined and is two times continuously differentiable in a bounded domain
${\mathcal D}_n$ of ${\mathbb R}^n$ with boundary $b({\mathcal D}_n)$. Suppose further that
the solutions of $F_n(x) = {\mathit \Theta}_n$
are not located on $b({\mathcal D}_n)$, and that they
are simple (that the determinant of the Jacobian of $F_n$ at these solutions is non-zero).
Then the {\em topological degree of $F_n$ at ${\mathit \Theta}_n$ relative to}
 ${\mathcal D}_n$ is denoted by
 ${\rm deg}[ F_n , {\mathcal D}_n,$ $ {{\mathit \Theta}_n }] $
and  can be defined by the following relation:
\begin{equation*}
 {\rm deg}[ F_n , {\mathcal D}_n , {{\mathit \Theta}_n} ]
=\displaystyle\sum_{{\scriptstyle x \in F_n^{-1} ( {{\mathit \Theta}_n} )} }
  {\rm sgn}\,{\rm det}\,J_{F_{n}} (x),\label{eq:deg}
\end{equation*}
where ${\rm det}\,J_{F_{n}} (x)$ denotes the determinant of the Jacobian matrix
and ${\rm sgn}$ defines the three-valued sign function.
The above definition can be generalized when $F_n$ is only continuous
{\cite{OrtegaR2000}}.

It is evident that, since  ${\rm deg}[ F_n , {\mathcal D}_n,$ $ {{\mathit \Theta}_n }] $
is equal to the number of simple solutions of $F_n(x) = {\mathit \Theta}_n$ which
give positive determinant of the Jacobian matrix, minus the number of simple solutions which give negative
determinant of the Jacobian matrix, then the total number $N^s$ of simple solutions of
$F_n(x) = {\mathit \Theta }_n$ can be obtained by the value of
${\rm deg}[ F_n , {\mathcal D}_n,$ $ {\mathit \Theta}_n] $ if all these solutions
have the same sign of the determinant of the Jacobian matrix. Thus, Picard considered
the following extensions of the function $F_n$ and 
the domain ${\mathcal D}_n$~\cite{Picard1892,Picard1922}:
\begin{equation}\label{eq:pfn}
F_{n+1}=(f_1, f_2, \ldots,f_n,f_{n+1})\colon {\mathcal D}_{n+1} \subset
{\mathbb R}^{n+1}\to {\mathbb R}^{n+1}, 
\end{equation}
where\
$f_{n+1} = y\,\,{\rm det}\,J_{F_{n}}$,\,\, ${\mathbb R}^{n+1} : x_1, x_2,\ldots,x_n,y$,\, and
${\mathcal D}_{n+1}$ is the direct product of the domain ${\mathcal D}_{n}$
with an arbitrary interval of the real $y$-axis containing the point
$y=0$. Then the solutions of the following system of equations:
\begin{equation*}
\begin{array}{l}
f_{i}(x_1,x_2,\ldots,x_n) = 0,\quad i=1, 2,\ldots,n,\\[0.1cm]
y\,\,{\rm det}\,J_{F_{n}}(x_1,x_2,\ldots,x_n)=0,\label{eq:psys}
\end{array}
\end{equation*}
are the same simple solutions of $F_n(x) = {\mathit \Theta }_n$ provided that $y=0$.
Obviously, the determinant of the Jacobian matrix obtained for the function (\ref{eq:pfn})
is equal to $({\rm det}\,{J_{F_{n}}}(x))^2$ which is always positive.
Thus, the total number $N^s$ of the solutions of the system 
$F_n(x) = {\mathit \Theta }_n$ can be obtained by the following value of the topological degree:
\begin{equation*}
N^s = {\rm deg}[ F_{n+1}, {\mathcal D}_{n+1}, {\mathit \Theta }_{n+1} ].
\end{equation*}

For example, in the one dimensional case, using the above Picard's extensions it is proved that the total number of 
simple solutions $N^s $ of the equation $f(x)=0$, where $f\colon (a,b) \subset $ 
${\mathbb R}$ $\to $ ${\mathbb R}$ is twice continuously differentiable in a predetermined interval $(a,b)$, is given by the following relation~\cite{Picard1892,Picard1922}:
\begin{equation}\label{eq:NR}
{N^s} = - \frac{1}{\pi}\left[ \varepsilon
     \int_a^b \frac{f(x)\,f''(x)-{f'}^2(x)}{f^2(x) + {\varepsilon}^2 {f'}^2(x)}\, dx +
 \arctan\left(\frac{\varepsilon f'(b)}{f(b)} \right)
 - \arctan\left(\frac{\varepsilon f'(a)}{f(a)} \right) \right],
 \end{equation}
where $\varepsilon$ is a small positive constant. Note that ${N^s}$ was shown to be independent of
the value of $\varepsilon$. 
Also, the above approach can be applied for computing the number of multiple solutions.
Obviously, the total number  $N^e$ of the extrema of $f\in C^3$\,
i.e.\ $x \in (a,b)$ such that $f'(x)=0$ can be obtained using the above formula~(\ref{eq:NR}).
For details of the topological degree we refer the interested reader to the books~\cite{Cronin1964,Lloyd1978,OReganCC2006,OutereloR2009,OrtegaR2000,Sikorski2001}. 
Details of the computation of the value of the topological degree and its usefulness as well as  
some applications and issues related to the number of zeros can be found for example 
in~\cite{BergaminBV2002,EmirisMV1999,KavvadiasV1996,Kearfott1979,MourrainPTV2006,MourrainVY2002,PlagianakosNV2001,PolymilisSSTV2003,Stenger1975,Stynes1979,Vrahatis1995,Vrahatis1988a,Vrahatis1989,VrahatisI1986,VrahatisSTB1993}. 

Article \cite{Franek2015} also discusses utilizing the topological degree to determine the existence of robust solutions to a SNE.

\section{Root finding methods}\label{rootfindingmethods}
This section discusses root finding methods which have been utilized to search for solutions to  SNEs. Specifically, Section \ref{newton} - Section \ref{lm} discuss methods which have been utilized to search for a single solution to a SNE, and 
Section \ref{deflation} discusses \textit{deflation techniques}\/ which can be used to compute further solutions. Some of the root finding methods introduced in this section are discussed in more detail than others, and after we introduce fundamental root finding methods, we place a particular emphasis on promising methods which do not appear to be widely discussed in recent literature. We also briefly introduce some of the more well known root finding methods such as \textit{Quasi-Newton methods}, and we cite relevant literature for the interested reader to refer to.

Although many of the root finding methods described in this section are guaranteed to converge to a solution when applied to SNEs that satisfy certain conditions such as monotonicity (Eq. (\ref{monotonicity})) and Lipschitz continuity (Eq. (\ref{l-continuity})), none of the methods described in this section are guaranteed to converge to a solution on all classes of SNEs. For example, the root finding method proposed in article \cite{Ullah2021} was proven to converge on systems of monotone nonlinear equations which satisfy additional regularity conditions, but the proposed method is not guaranteed to converge on all SNEs. 



Many of the root finding methods discussed in this section are specialized at solving particular classes of SNEs.
For example, Section \ref{TF-CH} discusses a \textit{tensor-free Chebyshev-Halley method} designed to solve SNEs that are large, sparse, and which may have ill-conditioned or singular Jacobian matrices at a solution \cite{eustaquio2018new}. 




\subsection{Classical Newton's and Broyden's methods for finding a single solution}\label{newton}
{\em Newton's method}~\cite{DennisS1996,kelley2003,OrtegaR2000,rheinboldt1998} for finding roots 
is a well known approach which can be utilized to find a single solution to a SNE. 
Starting at a ``good'' initial guess or approximate solution $x^{0}=(x_1^0, x_2^0, \ldots, x_n^0)^{\top}$ of a system of $n$ nonlinear real equations in $n$ real unknowns, when ${F}_n(x)$ is Lipschitz continuous and the Jacobian matrix $J_{{F}_n}(x)$ of ${F}_n(x)$ is available, Newton's method can be used to iteratively find elements of a sequence converging towards a true solution 
$x^* =(x_1^*, x_2^*, \ldots, x_n^*)^{\top}$ with a quadratic rate of convergence. Newton's method attempts to calculate a new approximate solution $x^{k+1}$ from $x^k$ via
\begin{equation*}\label{eq:5}
x^{k+1}=x^k-J_{{F}_{n}}\bigl(x^k\bigr)^{-1}\,{F}_n\bigl(x^k\bigr), \kern0.5cm k=0, 1, 2, \ldots
\end{equation*}
Alternatively, instead of calculating the inverse of the Jacobian matrix, one often 
can solve the system of linear equations: 
\begin{equation}\label{eq:6}
J_{{F}_{n}}\bigl(x^k\bigr)\,s^k=-{F}_n\bigl(x^k\bigr), \kern0.5cm k=0, 1, 2, \ldots
\end{equation}
to find the {\em Newton step}\, $s^k = x^{k+1}-x^k$. 
Thus, Newton's method using a initial guess~$x^0$ for each iteration $k=0, 1, 2, \ldots$ performs the following steps:\vspace*{0.1cm}
\begin{itemize}
\leftskip=0.4cm
\setlength{\itemsep}{-3pt}
\item[ (a) ] {\bf Solve:} the system of linear equation \,$ J_{{F}_{n}}\bigl(x^k\bigr) s^k = - F_n\bigl(x^k\bigr)$ \ for $s^k$.\\[-0.1cm]
\item[ (b) ] {\bf Set:} \,$x^{k+1} = x^{k} + s^k$.
\end{itemize}

\noindent
In many cases, Step (b) is given by $x^{k+1} = x^{k} + \lambda s^k$ where $\lambda$ is selected to guarantee decrease in $\|F_n\|$~(see \cite{kelley2003}).

Many expansions upon Newton's method for finding roots have been introduced in literature, and most of these expansions are focused on accelerating the rate of convergence. For example, articles \cite{frontini2004,darvishi2007b,srivastava2021,ASLAMNOOR2009, Ben-Israel1965, Yudin2021} present new techniques based upon Newton's method for finding roots and quadrature rules to accelerate convergence. Newton's method can also be augmented with higher order information 
to accelerate convergence towards an approximation of a true solution $x^*$. 
Overall, Newton's method for finding roots seeks to approximate the solution of a nonlinear system of equations by solving iteratively a sequence of systems of linear equations. In cases where the Jacobian matrix is singular, we refer the interested reader to articles \cite{BENISRAEL1966243, HUESO200977}. Convergence results for SNEs which have a Jacobian with a constant rank in the vicinity of a solution are presented in articles~\cite{Gatilov2014,xu2007a}.
Dimension-reducing modifications of Newton's method that are able to tackle efficiently and effectively, among others, almost linear systems can be found in~\cite{grapsa1989,grapsa1990,grapsa1995,grapsa1996,grapsaVB1990}.

In a similar approach, {\em Broyden's method}\/ \cite{broyden1965,DennisS1996,kelley2003,OrtegaR2000,rheinboldt1998} using an initial guess~$x^0$ 
and an initial matrix $B_0$ such that ${\rm det}\,B_0 \neq 0$,
for each iteration $k=0, 1, 2, \ldots$ performs the following steps:\vspace*{0.1cm}
\begin{itemize}
\leftskip=0.4cm
\setlength{\itemsep}{-3pt}
\item[ (a) ] {\bf Solve:} the system of linear equations \,$ B_k s^k = - F_n\bigl(x^k\bigr)$ \ for $s^k$.\\[-0.1cm]
\item[ (b) ] {\bf Set:} \,$x^{k+1} = x^{k} + s^k$.\\[-0.1cm]
\item[ (c) ] {\bf Set:} \,$y^k = F_n\bigl(x^{k+1}\bigr) - F_n\bigl(x^k\bigr)$.\\[-0.1cm]
\item[ (d) ] {\bf Set:} \,
$\displaystyle B_{k+1}= B_{k} + \displaystyle\frac{1}{\bigl(s^k\bigr)^{\top} s^k}\, 
\bigl(y^{k} - B_k s^k \bigr) \bigl(s^k\bigr)^{\top}.
$
\end{itemize}

\noindent
A good ``choice'' of $B_0$ is $B_0 = J_{{F}_{n}}\bigl(x^0\bigr)$. Obviously, by avoiding this choice Broyden's method does not require the computation of the Jacobian matrix. The effectiveness of Broyden's method and Newton's method on different classes of SNEs has been studied, for example, article \cite{GomesRuggiero1992} compares the effectiveness of Newton's method, Broyden's method, and other methods at solving sparse SNEs. 

There are many techniques that can be utilized to solve the system of linear equations produced by each iteration of Newton's or Broyden's method for finding roots. Two widely used methods for solving a system of linear equations are described in the following sections, and we refer the interested reader to \cite{OrtegaR2000} for more methods and details regarding these methods. For a visualization of the basins of convergence for Newton's and Broyden's methods, we refer the interested reader to article \cite{epitropakis2011}.


\subsubsection{Generalized Minimum Residual method (GMRES)}\phantom{}\vspace*{0.1cm}

\noindent
Consider a system of linear equations of the form $Ax=b$. The GMRES method \cite{Saad1986GMRES} is an iterative method which at the $k-$th iteration uses the Arnoldi iteration \cite{Arnoldi1951ThePO} to find an approximate solution $x^k$ in the $k-$th Krylov subspace $\mathcal{K}_k$ which minimizes the residual $\|b-Ax^k\|_2$. The $k$-th Krylov subspace of the GMRES method is
\begin{align*}
\mathcal{K}_k=\mathcal{K}_k(A,r^0)&={\rm span}\{r^0,Ar^0,A^2r^0,\ldots,A^{k-1}r^0\},
\end{align*}
where $r^0=b-Ax^0$ is the residual at the initial point $x^0$.


GMRES assumes that the matrix $A$ is invertible (and hence square) and that $b$ is normalized such that $\|b\|_2=1$. It is worth mentioning that because $A$ is assumed to be invertible, that  this technique in its original form can not be utilized with Newton's method to find an approximate solution to a SNE that has more equations than unknowns $(m >n)$.

\subsubsection{Successive Overrelaxation method (SOR)}\phantom{}\vspace*{0.1cm}

\noindent
The SOR method is another iterative method for solving a system of linear equations of the form $Ax=b$. If we assume that the diagonal elements $a_{ii}$ of $A$ are all non-zero, at the $k-$th iteration we can utilize the SOR iteration to find a new approximate solution $x^{k+1}$ to $Ax=b$ by the following update:
\[
x^{k+1}=x^k-\omega (D-\omega L)^{-1} (Ax^k-b),
\]
where 
$\omega>1$
is a relaxation parameter, and $A=D-L-U$ where $D,L$, and $U$ are diagonal, strictly lower triangular, and strictly upper triangular matrices respectively. Here, $L$ and $U$ have zero diagonal elements, and the assumption that the diagonal elements of $A$ are nonzero ensures that $(D-L)^{-1}$ exists \cite{OrtegaR2000}. The Gauss-Seidel iteration is a special case of the SOR where $\omega=1$. 

Though the SOR method and the corresponding Jacobi method \cite{OrtegaR2000} can be utilized to solve the series of linear systems produced by Newton's method, the SOR and Jacobi methods can also be extended to find a single solution to a SNE.

\subsection{Solving a SNE using generalizations of iterative methods for linear systems}

The well known and widely used {\em Gauss-Seidel}\/ iterative method for solving a linear system
of equations of the form  $A x=b$ can
be generalized for solving SNEs. 
Thus, if 
${F}_n=(f_1,f_2,\ldots, f_n): {\mathcal D} \subset {\mathbb R}^n
\to {\mathbb R}^n$, then the basic step of the {\em nonlinear
Gauss-Seidel (NGS)}\, iteration is to solve (in analogy to linear case),
the $i-$th nonlinear one-dimensional equation:
 \begin{equation}\label{NNGS}
  f_i(x_1^{k+1},\ldots,x_{i-1}^{k+1},x_i,x_{i+1}^k,\ldots,x_n^k)=0,
 \quad k=0,1,\ldots, \quad i=1, 2, \ldots,n.
 \end{equation}
for $x_i$, and to set $x_i^{k+1}=x_i$.
To obtain $x^{k+1}$ from
$x^k$, we solve successively the $n$ one-dimensional nonlinear
equations~(\ref{NNGS}) for $i=1,2,\ldots,n$. If relaxation
parameters $\omega_k$ are used, we may set $x_i^{k+1}=x_i^k
+\omega_k (x_i -x_i^k)$ and the corresponding method is called 
{\em Nonlinear Successive Overrelaxation}\/ (NSOR) method. 
In an analogous way, the $k-$th step of the {\em Nonlinear Jacobi}\/ (NJ)
iterative scheme consists of solving the $i-$th nonlinear one-dimensional equation:
 \begin{equation}\label{NNJAC}
  f_i(x_1^{k},\ldots,x_{i-1}^{k},x_i,x_{i+1}^k,\ldots,x_n^k)=0,
 \quad k=0,1,\ldots, \quad i=1, 2, \ldots,n.
 \end{equation}
for $x_i$ and setting $x_i^{k+1} = x_i$, for $i=1,2,\ldots,n$.
The main characteristic of the nonlinear Jacobi approach is that it can be easily parallelized.

In contrast to the linear case, in general, the analytic solutions
of Eqs.~(\ref{NNGS}) and~(\ref{NNJAC}) are not available and an
one-dimensional rootfinding method must be applied that
terminates after a suitable number of steps. Any type of
one-dimensional methods can be used leading to a large variety of
combined methods. On the other hand if many steps of these
one-dimensional methods are applied the whole procedure becomes
cumbersome and thus in practice, in many cases, a few steps or at least one step of these
methods is applied. In this case, for example, if we apply the {\em one-dimensional 
Newton's method}, the derivative-free {\em one-dimensional secant method}\/ or the derivative-free {\em one-dimensional Steffensen method}\/ to Eqs.~(\ref{NNGS}) and~(\ref{NNJAC}), we obtain respectively the following methods for solving SNEs~\cite{OrtegaR2000}:

\vspace*{0.2cm}
\noindent
(a) \ the {\em one-step SOR-Newton method}:
\begin{equation*}\label{onestepSORn}
x_i^{k+1} = x_i^k -  \omega_{ k}\,
{\frac{f_i ( x^{k,i} )}
  { \partial_i f_i ( x^{k,i}) }}, 
 \quad k=0,1,\ldots, \quad i=1, 2, \ldots,n.      
\end{equation*}
where $ x^{k,i} = (x_1^{k+1}, x_2^{k+1},\ldots, x_{i-1}^{k+1}, x_i^k ,
x_{i+1}^k, \ldots , x_n^k)^{\top}$ and
where $\partial_i f_i(x)$ denotes
the partial derivative of the function $f_i(x)$ with respect to the variable $x_i$.

\vspace*{0.2cm}
\noindent
(b) \ the {\em one-step Jacobi Newton method}:
\begin{equation*}\label{onestepJACn}
x_i^{k+1} = x_i^k -  \omega_{ k}\,
{\frac{f_i ( x^{k} )}
  { \partial_i f_i ( x^{k}) }}, 
 \quad k=0,1,\ldots, \quad i=1, 2, \ldots,n.
\end{equation*}

\vspace*{0.2cm}
\noindent
(c) \ the derivative-free {\em one-step SOR secant method}:
\begin{equation*}\label{onestepSORSec}
x_i^{k+1} = x_i^k -  \omega_{ k}
{\frac{ x_i^{k} - x_i^{k-1}}
  { f_i ( x^{k,i}) - f_i \bigl( x^{k,i} +( x_i^{k-1} - x_i^k)\>\!e^i\,\bigr) }}
f_i \bigl( x^{k,i} \bigr),
\quad k=0,1,\ldots, \quad i=1, 2,\ldots,n,
\end{equation*}
where
$e^i$ indicates the $i$-th column of the identity matrix~$I_n$.

\vspace*{0.2cm}
\noindent
(d) \ the derivative-free {\em one-step Jacobi secant method}:
\begin{equation*}\label{onestepSORSec_b}
x_i^{k+1} = x_i^k -  \omega_{ k}
{\frac{ x_i^{k} - x_i^{k-1}}
  { f_i ( x^{k}) - f_i \bigl( x^{k} +( x_i^{k-1} - x_i^k)\>\! e^i\,\bigr) }}
f_i ( x^{k} ),
\quad k=0,1,\ldots, \quad i=1, 2,\ldots,n.
\end{equation*}

\vspace*{0.2cm}
\noindent
(e) \ the derivative-free {\em one-step SOR Steffensen method}:
\begin{equation*}\label{onestepSORSteff}
x_i^{k+1} = x_i^k -  \omega_{ k}
{\frac{f_i \bigl( x^{k,i} \bigr)^2}
  { f_i ( x^{k,i}) - f_i \bigl( x^{k,i} - f_i ( x^{k,i})\>\!  e^i\,\bigr) }},
\quad k=0,1,\ldots, \quad i=1, 2,\ldots,n.
\end{equation*}

\vspace*{0.2cm}
\noindent
(f) \ the derivative-free {\em one-step Jacobi Steffensen method}:
\begin{equation}\label{onestepJACSteff}
x_i^{k+1} = x_i^k -  \omega_{ k}
{\frac{f_i \bigl( x^{k} \bigr)^2}
  { f_i ( x^{k}) - f_i \bigl( x^{k} - f_i ( x^{k})\>\! e^i\,\bigr) }},
\quad k=0,1,\ldots, \quad i=1, 2,\ldots,n.
\end{equation}

Furthermore, since in many problems of practical interest the functions
values are known only imprecisely, the traditional and widely applied {\em one-dimensional bisection method}
can be used for a suitable number of steps for computing solutions of Eqs.~(\ref{NNGS}) and~(\ref{NNJAC}). 
Specifically, for the computation of a zero of a continuous function 
$f\colon [a,b] \subset $ ${\mathbb R}$ $\to $ ${\mathbb R}$
the one-dimensional bisection method has be given in~\cite{Vrahatis1988a} by
the following sequence:
\begin{equation}\label{eq:bis}
x^{k+1} = x^k + {\rm sgn} f(x^0) \, {\rm sgn} f(x^k) \,
            (b-a) / 2 ^{k+1},\qquad x^0 = a, \qquad
            k=0,1,\ldots\,.
\end{equation} 
Similarly, instead of the above sequence we can also use the following one:
\begin{equation}\label{eq:biss}
x^{k+1} = x^k - {\rm sgn} f(x^0) \, {\rm sgn} f(x^k) \,
            (b-a) / 2 ^{k+1},\qquad x^0 = b, \qquad
            k=0,1,\ldots\,.
\end{equation}
Obviously, the sequences (\ref{eq:bis}) and (\ref{eq:biss}) converge to 
a root $r \in (a,b) $ if for some $x^k,\,\, k = 1,2,\ldots\,$ it holds that
${\rm sgn} f(x^0) \, {\rm sgn} f(x^k) = -1.$ 
Also, 
the number of iterations $\nu$, that 
are required in obtaining an approximate root $r^{*}$ 
such that $|r - r^{*} | \leqslant \varepsilon $ for some $\varepsilon \in (0,1) $
is given by:
\begin{equation}\label{eq:nbis}
\nu = \bigl\lceil {\log }_2 \bigl((b-a) \, {\varepsilon }^{-1}\bigr) \bigr\rceil.
\end{equation}
The one-dimensional bisection method always
converges within the given interval $(a, b)$ and is a globally
convergent method. Furthermore, it has a great advantage since it is
worst-case optimal; i.e., it possesses asymptotically the best
possible rate of convergence in the worst case~\cite{sikorski1982,Sikorski2001}. 
Thus, it is guaranteed to converge to an approximate root with a predetermined accuracy using the predefined number of iterations given by Eq.~(\ref{eq:nbis}). Notice that no other method has this
characteristic. 
This method actually requires only the signs of function values (and the gradient values for the optimization case) to be correct. Thus, it requires only one bit of information, namely the sign of a function value and consequently it can be applied to problems with imprecise
function values.
Also, this method can be generalized to tackle SNEs and optimization problems. In~\cite{Vrahatis1988a} a straightforward generalization of the bisection method, named {\em characteristic bisection method}\/ has been presented.
For a few details of the bisection and the generalized bisection methods and for some applications we refer the interested reader to
%
\cite{KavvadiasMV1999,
KavvadiasMV2005,
KavvadiasV1996,
MourrainVY2002,
Vrahatis1986,
Vrahatis1995,
Vrahatis1988b,
Vrahatis1988a,
VrahatisAM1996,
VrahatisI1986,
VrahatisIB1997,
ZottouKMV2018}.

The main characteristic of all the above briefly described methods in this section is that they can tackle $n$-dimensional SNEs using only one-dimensional rootfinding methods. 
This issue is very important in cases where the dimension $n$ is large.
Also, the corresponding algorithms are very simple to implement with a few lines of code.
Notice that, the convergence
properties of all the above methods are well studied and analyzed
(see for example~\cite{OrtegaR2000}).
On the other hand, although the nonlinear iterative rootfinding methods have been
extensively studied, the optimization case has not been thoroughly studied and analyzed.
Details and some applications of the above described methods can be found for example in
%
\cite{AlexandropoulosPV2020,
BoutsinasV2001,
VrahatisALM2000,
VrahatisAM1996,
VrahatisMP2000}.

\subsection{Chebyshev-Halley methods}\label{halleyMethod}
Although Newton's method only has a quadratic rate of convergence, \textit{Halley's method} possesses a cubic rate of convergence. {Halley's method}, also known as \textit{the method of tangent hyperbolas}, utilizes information from the tensor of second derivatives of ${F}_n(x)$ to accelerate convergence towards a local minimizer $x^*$ of ${F}_n(x)$ from an initial guess $x^0$. {Halley's method} attempts to calculate a new approximate solution $x^{k+1}$ to a SNE from $x^k$ using the following scheme: 
\[
x^{k+1}=x^k-\Bigl\{I+\frac{1}{2}L_{{F}_n}(x^k)\,\bigl[I-\alpha L_{{F}_n}(x^k)\bigr]^{-1}\Bigr\}\,{F}_n'(x^k)^{-1}{F}_n(x^k),
\]
where $\alpha$ is a real parameter, and $L_{{F}_n}(x)={F}_n'(x)^{-1} {F}_n''(x) {F}_n'(x)^{-1}{F}_n(x)$ is the degree of logarithmic convexity \cite{HERNANDEZ199927,hernandez1993family,gutierrez_hernandez_1997}. Therefore, {Halley's method} is applicable to cases where ${F}_n''(x)$ can be computed. 

{\em Chebyshev's method}\/ refers to the case where $\alpha=0$, the {\em classical Halley's method}\/ refers to the case where $\alpha=\frac{1}{2}$, and the {\em super-Halley method}\/ refers to the case where $\alpha=1$. Details about \textit{Chebyshev-like methods} for solving SNEs can be found in \cite{babajee2010}.


\subsection{Tensor methods utilizing higher order derivatives}
There exists a class of \textit{tensor methods} which utilizes information from higher order derivatives to accelerate convergence to a single solution of a SNE. In \cite{STEIHAUG2013230}, Steihaug and Suleiman utilize the model: 
\begin{equation}\label{eq:4}
    \|M^k(d)\|_2=\Bigl\|{F}_n(x^k)+J_{{F}_n}(x^k) d+\frac{1}{2}T^k d d\,\Bigr\|_2 \leqslant \eta^k \|{F}_n(x^k)\|_2.
\end{equation} 
which seeks to determine a step $d^k$ at each iteration. In this model, $T^k={F}_n''(x^k)$ is the tensor of second derivatives of ${F}_n(x^k)$, and $\eta^k \in [0,1)$. 
Steihaug and Suleiman proved that any method that can be used to find $d^k$ in Eq. (\ref{eq:4}) while satisfying $\|d^k\|_2=O(\|{F}_n(x^k)\|_2)$ is locally convergent. Furthermore, Steihaug and Suleiman showed that the rate of convergence is at least:
\begin{itemize}
\setlength{\itemsep}{4pt}
    \item[(a)] Q-super-linear when $\eta^k \rightarrow 0$.
    \item[(b)] Q-quadratic when $\eta^k=O(\|{F}_n(x^k)\|_2)$.
    \item[(c)] Q-cubic when $\eta^k=O(\|{F}_n(x^k)\|_2^2)$.
    \item[(d)] Q-order $\min \bigl\{\hat{p},\, 3 \bigr\}$ when $\eta^k=O(\|{F}_n(x^k)\|_2^{\hat{p}-1}),\ \ 1<\hat{p}$.
\end{itemize}
Furthermore, in order to solve Eq. (\ref{eq:4}), Steihaug and Suleiman in \cite{STEIHAUG2013230} introduce a class of {\em inexact Chebyshev-Halley methods}\/ which under some assumptions are locally convergent satisfying Eq.~(\ref{eq:4}) and $\|d^k\|_2=O(\|{F}_n(x^k)\|_2)$.

\subsection{Tensor-free Chebyshev-Halley method}\label{TF-CH}
Instead of calculating the tensor term $T^k d d$ as performed in Eq.~(\ref{eq:4}), Eustaquio {\em et al.}\/ in \cite{eustaquio2018new} avoid calculating the tensor term altogether by introducing a general \textit{tensor-free Chebyshev-Halley method} for solving SNEs. Therefore, though Eq.~(\ref{eq:4}) requires calculating ${F}_n''(x)$, the method of Eustaquio {\em et al.} can be utilized in cases where ${F}_n''(x)$ does not exist. Also, this method does not require the inequality in Eq.~(\ref{eq:4}) to be satisfied. The general framework for the {\em inexact tensor-free Chebyshev-Halley class}\/ is the following:

\begin{enumerate}
\setlength{\itemsep}{4pt}
    \item Given the SNE ${F}_n(x) = {\mathit \Theta}_{n}$, the machine precision $\varepsilon_M$, and $k=0$, select $\alpha \in {\mathbb R}$, a step length $h>0$, an initial point $x^0 \in {\mathbb R}^n$, a forcing term limit $\Tilde{\eta} \in (0,1)$, and a mapping $C : {\mathbb R}^n \rightarrow {\mathbb R}^{n \times n}$.
    \item \textbf{while } ${F}_n(x^k)\neq {\mathit \Theta}_{n}$ \textbf{ do}
    \begin{enumerate}
    \setlength{\itemsep}{4pt}
        \item Select forcing term tolerances $\eta_{1}^k \in [0,\Tilde{\eta})$ and $\eta_{2}^k \in [0,\Tilde{\eta})$.
        \item Compute $d_1$ such that $\bigl\|J_{{F}_n}(x^k)d_1 + {F}_n(x^k)\bigr\|_2 \leqslant \eta_{1}^k \bigl\|{F}_n(x^k)\bigr\|_2$.
        \item Compute $d_2$ such that $\bigl\|\bigl(J_{{F}_n}(x^k)+\alpha\, C(x^k)\bigr) d_2 + \frac{1}{2}C(x^k) d_1\bigr\|_2 \leqslant \eta_{2}^k \bigl\|\frac{1}{2}C(x^k) d_1\bigr\|_2$.
        \item Set $x^{k+1}=x^k+d_1+d_2$.
        \item Set $k=k+1$.
    \end{enumerate}
    \item \textbf{end while}
\end{enumerate}
where the authors utilized the mapping $C(x^k)= \frac{1}{h} \bigl( J_{{F}_n} (x^k+h d_1)-J_{{F}_n}(x^k) \bigr)$ with step length $h=\sqrt{\varepsilon_M}\, \|x^k\|_2 \big/ \|d_1\|_2 \in (\varepsilon_M,0.5)$ and the forcing terms $\eta_{1}^k=\min \bigl\{ 10^{-8},\, \|{F}_n(x^k)\|_2^2 \bigr\} \in (2\varepsilon_M, 10^{-8}]$ and $\eta_{2}^k=\min \bigl\{ 10^{-8},\, \|{F}_n(x^k)\|_2 \bigr\} \in (2\varepsilon_M, 10^{-8}]$ to control the level of accuracy of the approximate solution to the SNE. In comparison to Eq.~(\ref{eq:4}), the mapping $C(x^k)$ replaces the tensor term $T^k d$ while still preserving a third-order rate of convergence. As shown in the algorithm above, each iteration only requires approximately solving two linear systems. Eustaquio {\em et al.}\/ utilized Saad and Schultz's  Generalized Minimum Residual method (GMRES) \cite{Saad1986GMRES} to solve the two linear systems and obtain the inexact Newton steps $d_{1}^k$ and $d_{2}^k$.  

The authors proved that any method belonging to their class of inexact tensor-free Chebyshev-Halley methods is locally convergent. Note that the system solved in step 2(b) is analogous to system (\ref{eq:6}) solved in Newton's method where $d_1=x^{k+1}-x^k$. 

Eustaquio {\em et al}. conducted a rigorous quantitative comparative analysis of their method against the \textit{Inexact Newton method} and the tensor methods utilizing higher order derivatives presented by Steihaug and Suleiman \cite{STEIHAUG2013230}. Eustaquio {\em et al.} illustrated that the class of tensor-free Chebyshev-Halley methods are capable of efficiently handling problems that have ill-conditioned or singular Jacobian matrices at the solution. 

Furthermore,
{\em Jarrat's method}\/ \cite{Argyros1994} is defined as a specific case of the {tensor-free Chebyshev-Halley method} where $\alpha=1$, $\eta_{1}^k=0$, $\eta_{2}^k=0$, $h={2}/{3}$, and the mapping 
$C(x^k)={h^{-1}} \bigl(J_{{F}_n} (x^k + h\,d_1)-J_{{F}_n}(x^k)\bigr)$. Jarrat's method has a convergence rate of four. Articles using Jarratt-like methods to solve SNEs can be found for example in \cite{khirallah2013solving,ahmad2015,zhanlav2021}.

\subsection{Tensor methods not utilizing higher order derivatives}\label{tensorWO-HOD}
In addition to the class of tensor methods which utilize a tensor of higher order derivatives, other tensor methods exist and have demonstrated effectiveness at solving large, sparse, and ill-formed SNEs with singular Jacobian matrices \cite{frank1984tensor,schnabel1986solving,bouaricha1992solving,bouaricha1994tensor,feng1997tensor,bouaricha1997algorithm,bouaricha1998tensor,bader2004performance,Bader2005}. 
In \cite{frank1984tensor} Frank and Schnabel introduced tensor strategies for solving SNEs based around solving the quadratic model:
\begin{equation}\label{eq:7}
    \min_{d \in {\mathbb R}^n} \|M^k(d)\|_2=\min_{d \in 
    {\mathbb R}^n}\Bigl\| {F}_n(x^k)+J_{{F}_n} (x^k) d +\frac{1}{2} T^k d d\,\Bigr\|_2,
\end{equation}
where $J_{{F}_n}(x)$ is the Jacobian matrix of ${F}_n(x)$, $d$ is the step size, and $T^k \in {\mathbb R}^{n\times n \times n}$ is a carefully chosen tensor. 

Expanding upon this formulation, in \cite{Bader2005}, Bader presents three Krylov-based methods for iteratively solving Eq, (\ref{eq:7}) to a specified tolerance. Bader's method selects $T^k$ such that the model interpolates $p \leqslant \sqrt{n}$ function values from its most recent history of iterates. By selecting $p=1$, Bader reduces the tensor model about $x^k$ to 
\begin{equation}\label{eq:3}
    \min_{d \in \mathcal{K}_m} \|M^k(d) \|_2=
    \min_{d \in \mathcal{K}_m} \Bigl\|{F}_n(x^k) + J_{{F}_n}(x^k) d + \frac{1}{2}a^k \bigl((s^k)^{\top} d\bigr)^2\Bigr\|_2,
\end{equation}
where ${\mathcal K}_m$ is a $m$-dimensional Krylov subspace, and where $a^k$, 
$s^k \in {\mathbb R}^n$ are given as follows:
\begin{align*}
    a^k &= 
    \frac{2\bigl({F}_n(x^{k-1})-{F}_n(x^k)-J_{{F}_{n}}(x^k)s^k)}{((s^k)^{\top}s^k\bigr)^2},\\[0.1cm]
    s^k &= x^{k-1}-x^k.
\end{align*}
The {\em linear Krylov subspace method}\/ finds an approximate solution $x_m$ to the linear system $Ax=b$ from an $m$-dimensional affine subspace $x^0+\mathcal{K}_m$ where 
\begin{align*}
\mathcal{K}_m(A,r^0)&={\rm span}\{r^0,Ar^0,A^2r^0,\ldots,A^{m-1}r^0\}, \\[0.1cm]
r^0&=b-Ax^0,
\end{align*}
and $r^0$ is the residual at the initial point $x^0$.

The three techniques presented by Bader in \cite{Bader2005} select $\mathcal{K}_m$ in different ways as a subroutine within the following algorithm:
\begin{enumerate}
\setlength{\itemsep}{4pt}
    \item Given the SNE ${F}_n(x)$, select an initial point $x_0$ and a maximum number of iterations $k_{max}$.
    \item For $k=0,1,2,\ldots,k_{max}$, do:
    \begin{enumerate}
    \setlength{\itemsep}{3pt}
        \item Select a forcing term tolerance $\eta^k \in [0,1)$.
        \item If $k=0$:
        \begin{enumerate}
        \setlength{\itemsep}{2pt}
            \item Calculate the Newton-GMRES \cite{Saad1986GMRES} step $d_N$ based on the tolerance $\eta^k$. 
            \item Proceed to step 2e. 
        \end{enumerate}
        \item Form the local tensor model (\ref{eq:7}).
        \item Calculate the approximate tensor step $d_T$ according to $\eta^k$ by solving one of the three methods presented for selecting $\mathcal{K}_m$.
        \item Set $x^{k+1}=x^k+\alpha d$ where a linesearch strategy using the directions $d_T$ and / or $d_N$ is used to select $d$ and $\alpha$.
        \item If $x^{k+1}$ is an acceptable approximate root of ${F}_n(x)$:
        \begin{enumerate}
        \setlength{\itemsep}{2pt}
            \item Stop.
        \end{enumerate}
    \end{enumerate}
\end{enumerate}


Bader's methods demonstrated effectiveness at solving large-scale SNEs, especially those which have ill-conditioned or singular Jacobians at the solution. 

\subsection{Quasi-Newton method for finding roots}
Although Newton's method for finding roots requires the Jacobian $J_{{F}_n}(x)$, if $J_{{F}_n}(x)$ is not available, \textit{Quasi-Newton methods} can be utilized. A Quasi-Newton method is any method that utilizes an approximation of $J_{{F}_n}(x)$ instead of the exact $J_{{F}_n}(x)$. A classification of Quasi-Newton methods for solving SNEs is presented in article \cite{Brezinski2004ACO}. Examples of Quasi-Newton methods include the methods discussed in articles  
%
\cite{Li1989, 
Mario1992}.

One of the most popular Quasi-Newton methods is the \textit{Broyden–Fletcher–Goldfarb–Shanno} (BFGS) \textit{method} \cite{broyden1970,fletcher1970,Goldfarb1970,shanno1970conditioning}. Popular extension of BFGS method is limited memory BFGS (L-BFGS) \cite{Liu1989}. Article \cite{Ullah2021} combines a memoryless 
variant of the BFGS method with a projection technique for solving SNEs, and the authors demonstrated the efficiency of the method by solving large SNEs with up to 5000 dimensions in the order of milliseconds or seconds, depending on the problem instance. Other Quasi-Newton methods include the \textit{Symmetric Rank 1} (SR1) \textit{method} and the \textit{Davidon–Fletcher–Powell} (DFP) \textit{method} among others. Quasi-Newton methods were utilized to find solutions to SNEs in articles 
%
\cite{birgin2003,
dauda2020, 
ek2020,
kamfa2020, 
leong2011,
luo2008,
mamat2016,
MARTINEZ2000,
perez2004}.

\subsection{Spectral methods}
\textit{Spectral methods} are another class of derivative-free methods which have been used to solve SNEs, and they are particularly well suited for solving large SNEs because they have a relatively low storage requirement \cite{aji2021}. Spectral methods often utilize spectral parameters and coefficients to help determine the search direction $d^k$ in the update step $x^{k+1}=x^k+\alpha^k d^k$. For example, article~\cite{aji2021} presents two \textit{derivative-free hybrid spectral methods} for solving SNEs, and proves that the proposed methods will converge to a solution when applied to SNEs that satisfy the conditions of monotonicity (Eq. (\ref{monotonicity})) and Lipschitz  continuity (Eq. (\ref{l-continuity})). Other articles which propose applying spectral methods for solving SNEs include \cite{papini2020,LaCruz2006}.


\subsection{Hermitian and skew-Hermitian splitting (HSS) based methods}
First introduced in \cite{bai2003} for solving linear systems of equations, \textit{Hermitian and skew-Hermitian splitting} (HSS) \textit{based methods} have been shown to be effective at solving sparse SNEs \cite{wu2013} and SNEs which can be decomposed into a linear part $Ax$ and a nonlinear part $\zeta (x)$ such that $F_m(x)=Ax-\zeta(x)$ \cite{bai2009}. When the linear part $Ax$ is dominant over the nonlinear part $\zeta (x)$, $F_m(x)$ can be referred to as a \textit{weakly nonlinear system}. By separating the linear and nonlinear elements of $F_m(x)$, certain specialized techniques can potentially be utilized to find solutions to $F_m(x)$ more quickly than by utilizing traditional root finding methods. Some of these techniques are derivative-free, including the derivative-free HSS based method introduced in \cite{amiri2019} which is guaranteed to converge to a solution on SNEs that satisfy certain conditions.

\subsection{Levenberg-Marquardt method}\label{lm}
The {\em Levenberg-Marquardt algorithm}~\cite{Levenberg1944,Marquardt1963} %
was designed to solve a nonlinear least squares problem that can be in general expressed as $\varphi(x) = \frac{1}{2}\sum_{i=1}^{m}[f_i(x)]^2$ where $\varphi: \mathbb{R}^n \rightarrow \mathbb{R}$.
The Levenberg-Marquardt algorithm is an iterative procedure which at each iteration calculates an updated solution using the rule $x^{k+1} = x^k + \lambda d$, where the search direction $d$\/ is found by solving equations of the form 
\[
\bigl({J}^{\top}_{{F}_m} {J}_{{F}_m} + \mu\,I_n\bigr) d = -{J}^{\top}_{{F}_m} {F}_m,
\]
where ${J}_{{F}_m}\in \mathbb{R}^{m\times n}$ is the Jacobian matrix of ${F}_m$, $\mu \in \mathbb{R}$, $I_n$ is the identity matrix, and ${F}_m: \mathbb{R}^n\rightarrow \mathbb{R}^m$. 
In article \cite{Fletcher1971}, Fletcher proposed the following modification  to make the solution scale invariant: 
\[
\bigl({J}^{\top}_{{F}_m} {J}_{{F}_m} + \mu\,{\rm diag} \{ {J}^{\top}_{{F}_m} {J}_{{F}_m} \} \bigr) d = -{J}^{\top}_{{F}_m} {F}_m,
\]
where 
the identity matrix $I_n$ is replaced with the diagonal matrix consisting of the diagonal elements of ${J}^{\top}_{{F}_m} {J}_{{F}_m}$.
Article \cite{kuzina2020} proposed to use the Levenberg-Marquardt method for solving a SNE modeling a port mooring structure. 
The Levenberg-Marquardt algorithm is widely used in many applications, including 
artificial neural network training (for example, see~article \cite{Rubio2021}).

\subsection{Deflation techniques for the computation of further solutions}\label{deflation}

\noindent
As illustrated above, there is a plethora of methods for obtaining a single solution of a system of~$n$ nonlinear real equations in $n$ real unknowns. Brown and Gearhart in \cite{BrownG1971} 
proposed  {\em deflation techniques}\/ for the computation of further solutions of a system of nonlinear equations. Specifically, these techniques proceed as follows: ``Once a solution of a system of nonlinear equations has been obtained a modified system is formed in such a way that it retains those solutions of the original system which remain to be computed except the solutions that has been already computed''. This procedure may be applied sequentially until all solutions of the original system are obtained.

For example, assume a system of nonlinear equations
${F}_n(x) = {\mathit \Theta}_{n}$, where 
${F}_n=(f_1,f_2,\ldots, f_n): {\mathcal D}_n \subset {\mathbb R}^n
\to {\mathbb R}^n$, then to deflate out the $p$ already computed roots $r_1$, $r_2, \ldots, r_p$
in order to compute additional roots, the following {\em norm deflated function} \cite{BrownG1971} is applied:
\[
\hat{f}_i (x) = \frac{1}{\prod_{j=1}^p\, \|x - r_j \|}\, {f}_i(x),
\kern0.5cm i=1, 2, \ldots, n ,
\]
or alternatively the following {\em inner product deflated function} \cite{BrownG1971} can be used:
\[
\tilde{f}_i (x) = \frac{1}{\prod_{j=1}^p\, \bigl\langle \nabla f_i (r_j),\, (x - r_j) \bigr\rangle}\, {f}_i(x),
\kern0.5cm i=1, 2, \ldots, n .
\]
The above inner product deflated function has proven to be useful in practice when Newton's method for root finding is used.
The deflation techniques can also be extended for the computation of multiple solutions. 
A study of the deflation techniques for the one dimensional case can be found in article \cite{Wilkinson1994}.
Also, additional techniques for the optimization case related to the above deflation techniques as well as some applications can be found for example
in the papers ~\cite{KalantonisPPRV2003,ParsopoulosPMV2001,ParsopoulosV2002,ParsopoulosV2004,ParsopoulosV2010,ParsopoulosV2021}.

\section{Symbolic computation methods}

Exact algorithmic methods for solving systems of nonlinear polynomial equations have been developed in the realm of \textit{Symbolic Computation}, also called \textit{Computational Algebra}, or \textit{Computer Algebra}. Broadly speaking, this particular subarea of Symbolic Computation can in fact be interpreted as a constructive version of some parts of the well-established mathematical 
theories of \textit{Commutative Algebra} and \textit{Algebraic Geometry}. We begin this section by a brief presentation (via examples) of the main Symbolic Computation methods for solving systems of nonlinear polynomial equations, namely \textit{resultants} and \textit{Gr{\"o}bner bases}.
We end this section by citing a number of books that discuss extensively all the developments in the area of Symbolic Computation exact methods for solving systems of nonlinear polynomial equations. 

\subsubsection{Resultants}\phantom{}\vspace*{0.1cm}

\noindent
Historically, the theory of resultants was developed with the aim to provide a systematic means of elimination of variables/unknowns from a system of nonlinear polynomial equations. 

The resultant of two univariate polynomials of degrees $m$, $n$ respectively, 
\[
p(x)=p_m x^m + \cdots + p_1 + p_0, \kern0.8cm q(x)=q_n x^n + \cdots + q_1 + q_0,
\]
is defined as the determinant of the $(n+m)\times(n+m)$ Sylvester matrix associated to $p(x)$ and $q(x)$, namely:
\[
    {\rm res}_x\bigl(p(x),q(x)\bigr) = 
    {\rm det} \left[
    \begin{array}{llll}
         a_m &  a_{m-1} &          & \\
          0  & a_m     & a_{m-1}  & \\
          0  &  0      & a_m      & a_{m-1} \\
          0  &  0      & 0        & a_m     \\
         b_n & b_{n-1} &          & \\
          0  & b_n     & b_{n-1}  & \\
          0  &  0      & b_n      & b_{n-1} \\
          0  &  0      & 0        & b_n \\ 
    \end{array}
    \right] .
\]
One of the main properties of the resultant is that it is equal to zero, if and only if the two polynomials have a common root.

The resultant of two bivariate polynomials $p=p(x,y)$,\kern0.1cm $q=q(x,y)$ with respect to the variable $x$, or with respect to the variable $y$, is the determinant of their associated Sylvester matrix, when $p$ and $q$ are considered as polynomials in $x$, or in $y$ respectively. 

Let us illustrate the concept of the resultant by a simple but instructive example. 
\begin{example}{\em
Consider the system of two polynomials $f_1(x_1,x_2) = 0, f_2(x_1,x_1) = 0$ in two variables $x_1,x_2$, given by:
\begin{equation}
    \begin{array}{l}
         f_1(x_1,x_2) = x_1^2+x_2^2-1, \\[0.1cm]
         f_2(x_1,x_2) = x_1-x_2-1. \\ 
    \end{array}
    \label{SymbolicComputation:Resultants:example}
\end{equation}
The resultant of $f_1,f_2$, with respect to $x_1$  is the determinant of their $3\times 3$ Sylvester matrix when considered as polynomials in $x_1$, namely, $f_1 = 1 \cdot x_1^2 + 0 \cdot x_1 + (x_2^2-1)$, $f_2 = 1 \cdot x_1 +(-x_2 - 1)$, 
$$
{\rm res}_{x_1}(f_1,f_2) = \det 
 \left[ \begin {array}{ccc} 1&0&{{\it x_2}}^{2}-1\\ \noalign{\medskip}1&-
{\it x_2}-1&0\\ \noalign{\medskip}0&1&-{\it x_2}-1\end {array} \right]
= 2x_2^2 + 2x_2 .
$$
By factorizing the resultant and setting it equal to zero, we obtain $x_2=0$ and $x_2= -1$. By substituting each of these two values of $x_2$ back into (\ref{SymbolicComputation:Resultants:example}), we obtain the corresponding values of $x_1$ as: $x_1=1$ and $x_1=0$. Therefore, we conclude that the system (\ref{SymbolicComputation:Resultants:example}) possesses the two solutions:
\[
    (x_1=1,\,\,x_2=0),\kern0.2cm (x_1=0,\,\,x_2=-1) , 
\]
which can easily be verified to be correct. In terms of geometric interpretation, the first equation $f_1(x_1,x_2)$ represent the unit circle and the second equation $f_1(x_1,x_2)$ represents a straight line, that intersects the unit circle. The two solutions of the system (\ref{SymbolicComputation:Resultants:example}) are the two points of intersection of the (red) unit circle and this (green) straight line in Figure 2 below. 
\begin{figure}[h]\label{fig4} 
\centering
 \includegraphics[width=7.5cm]{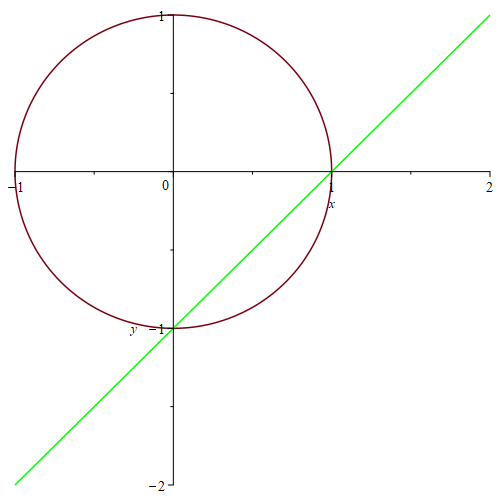}
  \caption{Geometric interpretation of System (\ref{SymbolicComputation:Resultants:example}) }
\end{figure}
}\end{example}

\subsubsection{Gr{\"o}bner bases}\phantom{}\vspace*{0.1cm}

\noindent
The theory of Gr{\"o}bner bases was developed as a multivariate analogue of the classical \textit{Gaussian elimination} for systems of linear equations. Again we illustrate the concept with an example and refer to the aforementioned books for additional details and technicalities. 
\begin{example}{\em
Consider the system of three polynomials $f_1(x_1,x_2,x_3) = 0$,\, $f_2(x_1,x_2,x_3) = 0$,\, $f_3(x_1,x_2,x_3) = 0$\, in three variables $x_1,x_2,x_3$, given by:
\begin{equation}
    \begin{array}{l}
         f_1(x_1,x_2,x_3) = x_1^{2}+x_2 x_3 -2, \\ [0.1cm]
         f_2(x_1,x_2,x_3) = x_1 x_3 +x_2^{2}-3, \\ [0.1cm]
         f_3(x_1,x_2,x_3) = x_1 x_2 +x_3^{2}-5. \\
    \end{array}
    \label{SymbolicComputation:Groebner:example}
\end{equation}
The lexicographical Gr{\"o}bner basis of the ideal generated by these three polynomials, with respect to the lexicographical ordering induced by $x_2 > x_1 > x_3$ is given by the three polynomials: 
\begin{equation}
    \begin{array}{l}
    8 x_2^{8}-60 x_2^{6}+142 x_2^{4}-172 x_2^{2}+1, \\[0.1cm]
88 x_2^{7}-680 x_2^{5}+1674 x_2^{3}-2081 x_2 +117 x_1, \\[0.1cm]
152 x_2^{7}-1132 x_2^{5}+2700 x_2^{3}-3403 x_2 +117 x_3. \\
    \end{array}
    \label{GB_plex}
\end{equation}
Now we notice that the first polynomial in (\ref{GB_plex}) depends only on $x_2$ and is of degree $8$. In addition, the second polynomial in (\ref{GB_plex}) depends linearly on $x_1$ and the third polynomial in (\ref{GB_plex}) depends linearly on $x_3$. This allows us to express $x_1$ and $x_3$ as polynomials in $x_2$. Therefore, the solution process starts by finding the 8 roots of the first polynomial in (\ref{GB_plex}) and for each one of those roots, we find the unique values of $x_1$ and $x_3$, given by the last two polynomials in (\ref{GB_plex}). Here are the eight complex roots (four real roots and two pairs of complex conjugate roots) of the first polynomial in (\ref{GB_plex})
$$
\begin{array}{l}
-2.16624183202470499 \\[0.1cm]
\kern0.22cm 2.16624183202470499 \\[0.1cm]
-0.07643337499454630 \\[0.1cm] 
\kern0.22cm 0.07643337499454630  \\[0.1cm]
-1.329681781358029 - 0.606033421098925 \,\mathrm{i} \\[0.1cm]
-1.329681781358029 + 0.606033421098925 \,\mathrm{i} \\[0.1cm]
\kern0.22cm 1.329681781358029 -0.606033421098925 \,\mathrm{i} \\[0.1cm]
\kern0.22cm 1.329681781358029 + 0.606033421098925 \,\mathrm{i} \\
\end{array}
$$
Note that the four real roots and the four complex roots come in pairs of the form $(r,-r)$, a consequence of the fact that the first polynomial in (\ref{GB_plex}) contains only even powers of $x_2$. The presence of only even powers of the indeterminate $x_2$ in the first polynomial in (\ref{GB_plex}) is captured by the fact that the order of the Galois group of this polynomial is smaller than the order of the Galois group of a more ``random/generic'' polynomial of degree $8$. 
}\end{example}
A very readable, self-contained and pedagogical introduction to  Gr{\"o}bner bases can be found in chapter 10 of \cite{GeddesCzaporLabahn:1992}.

\subsubsection{Symbolic computation software systems}\phantom{}\vspace*{0.1cm}

One of the major outcomes of the research area of Symbolic Computation (Computer Algebra) in the past few decades, 
is the advent of the so-called Computer Algebra Systems (CAS).
A CAS is defined as a piece of mathematical software that makes advanced functionalities available to the user, in a transparent manner. Such advanced functionalities include univariate and multivariate polynomial factorization, primality testing, integer factorization, root finding, numerical and exact integration, visualization, number theory, linear algebra, commutative algebra, commutative and non-commutative settings, tools for optimization, graph theory, group theory, coding theory, combinatorics, discrete mathematics and so forth.     
There is a number of commercial and free (open source) CAS available today. Some of the most well-known early CAS (such as Axiom and Macsyma) are largely deprecated today. We focus our attention on those CAS that currently feature a significant user base. We note that CAS are used by millions of research mathematicians, physicists, chemists, astronomers, engineers, practitioners and educators at academic, public, private and government institutions worldwide. In what follows, we provide a brief summary of the main commercial CAS in use today.

\begin{itemize}
    \item[(a)] Maple is the flagship Canadian product in the area of mathematical software. It is produced, maintained and distributed by the company Maplesoft, based in Waterloo, Ontario, Canada \url{https://www.maplesoft.com/}. Maple features an easy-to-learn underlying programming language as well as more than $150$ additional packages that significantly expand its core functionalities.   
    \item[(b)] Magma is the flagship Australian product in the area of mathematical software. It is produced, maintained and distributed by the University of Sydney \url{http://magma.maths.usyd.edu.au/}. It provides a mathematically rigorous environment for defining and working with structures such as groups, rings, fields, modules, algebras, schemes, curves, graphs, designs, codes and many others. Magma also supports a number of databases designed to aid computational research in those areas of mathematics which are algebraic in nature.
    \item[(c)] Mathematica is a flagship American product in the area of mathematical software. It is produced, maintained and distributed by the company Wolfram Research, based in Champaign, Illinois, United States \url{https://www.wolfram.com/}. Mathematica's stated purposes include injecting computational intelligence at every level, on every project by unifying algorithms, data, notebooks, linguistics and deployment—enabling powerful workflows across desktop, cloud, server and mobile.
    \item[(d)] Matlab is a flagship American product in the area of mathematical software.  \url{https://www.mathworks.com}. It is produced, maintained and distributed by the company MathWorks, based in California, United States. MATLAB's strong points include matrix manipulations, plotting of functions and data, implementation of algorithms, creation of user interfaces, and interfacing with programs written in other languages.
\end{itemize}

Some of the most popular non-commercial mathematical software packages include Sage, Singular/Plural, CoCoA, Macaulay 2 and so forth. Each one has their own strengths and areas of particular focuses. For more information on CAS in general, we refer the interested reader to the comprehensive book \cite{Wester:1999}.

\subsubsection{Symbolic computation books}\phantom{}\vspace*{0.1cm}

\noindent
In this section, we present a selection of books in Symbolic computation and related areas. While there is inevitably some overlap among these books, there are also different aspects of polynomial system solving and its applications discussed in each one of them.  

\begin{enumerate}
\setlength{\itemsep}{4pt}
    \item
The classic books \cite{GeddesCzaporLabahn:1992,Davenport:Siret:Tournier:1993,vzGathen:Gerhard:2013} are general references for Computer Algebra.

    \item
The series of books by Teo Mora
\cite{Teo_Mora:EMA_I_2003, Teo_Mora:EMA_II_2005, Teo_Mora:EMA_III_2015, Teo_Mora:EMA_IV_2016}
is an encyclopedic reference to the subject of polynomial system solving.

    \item
The theory of Gr{\"o}bner bases, see 
\cite{William_W_Adams_Philippe_Loustaunau:1994,%
Thomas_Becker_Volker_Weispfenning:1993,%
Ralf_Froberg:1997,%
Franz_Winkler:1996},
was initially developed by Bruno Buchberger 
and further extended by several other researchers.
The \textit{Buchberger algorithm} to compute Gr{\"o}bner bases 
of polynomial systems is implemented in every major 
Symbolic Computation software today. This includes commercial software, 
such as Maple, Magma and Mathematica and open source software, such as 
Singular \cite{Wolfram_Decker_Christoph_Lossen:2006}, 
Plural \cite{Wolfram_Decker_Gerhard_Pfister:2013}, 
CoCoA 
\cite{Martin_Kreuzer_Lorenzo_Robbiano:2000,%
Martin_Kreuzer_Lorenzo_Robbiano:2005,%
Martin_Kreuzer_Lorenzo_Robbiano:2008,%
Martin_Kreuzer_Lorenzo_Robbiano:2016}
and the Macaulay2 software system \cite{Macaulay2:Grayson:Stillman}.

    \item
The book \cite{Bernd_Sturmfels:2002} discusses aspects of systems of polynomial equations from the viewpoints of computational commutative algebra, discrete geometry, elimination theory, real geometry, as well as their applications in various domains such as partial differential equations, economics, probability, and statistics. 

    \item
The book \cite{Invariant_Theory_Sturmfels:1993} and its second edition \cite{Invariant_Theory_Sturmfels:2008}, describe useful algorithmic aspects of systems with symmetries, treated using \textit{Invariant Theory}.  
The book \cite{Karin_Gatermann:2000} is concerned with 
polynomial systems with symmetries, that arise in the 
context of dynamical systems. The french-language book \cite{Elkadi:Mourrain:2007} is devoted exclusively to polynomial system solving 
methods. The french-language book \cite{AECF:2017} contains a series of chapters on polynomial system solving.

    \item
The books \cite{Algebraic_Statistics_Pistone_Riccomagno_Wynn:2001} and \cite{Algebraic_Statistics_Sullivant:2018} describe the emerging area of 
\textit{Algebraic Statistics}, 
while
the book \cite{Hans_J_Stetter:2004} focuses on numerical aspects of
polynomial system solving.
On the other hand,
the book \cite{Alicia_Dickenstein_Ioannis_Z_Emiris:2005} focuses on 
theoretical aspects of polynomial system solving.
Furthermore,
the books trilogy 
\cite{Cox:2020:CBMS:134,CLO_UTM_2nd_ed:2005,CLO_UTM_4th_ed:2015} 
examine the foundation of polynomial system solving from the algebraic geometry standpoint as well as the applications standpoint.
The Galois group of a polynomial is inextricably linked with the more general \textit{Galois Theory}, 
we mention the three books \cite{GALOIS:Cox:2012,GALOIS:Stewart:2015,GALOIS:Tignol:2016}. 
\end{enumerate}

\section{Homotopy / Continuation methods}
\textit{Homotopy methods}, also referred to as \textit{continuation methods}, may be used for finding solutions to polynomial equations and systems of polynomial equations 
\cite{Allgower2003, 
Watson1990, Morgan2009}. 
Polynomial equations often arise in kinematics and robotics related problems 
\cite{Ji2003KinematicsAO,Lafmejani20151}; papers \cite{Chiang2014, mehta2016numerical} apply homotopy methods to solve systems of power flow equations; large-scale integrated circuit designs and protein-protein interaction equation are solved using homotopy method in \cite{Chiang2018}, \cite{Palancz2009} applies homotopy methods for solving SNEs arising in geodesy; \cite{Mehta2011} applies homotopy for finding string vacua, \cite{Jimenez-Islas2013} applies homotopy methods in chemical engineering.
Homotopy (or deformation) of a system of equations ${F}_n(x) = {\mathit \Theta}_{n} \equiv (0, 0, \ldots, 0)^{\top}$ (cf. Eq. (\ref{eq:1})) is a function $H_n$ such that $H_n(x, 1) = G_n(x)$, and $H_n(x, 0) = {F}_n(x)$, where the roots of $G_n(x)$ are known. For example, it is possible to choose a convex homotopy,  $H_n(x, \lambda) = \lambda G_n(x) + (1-\lambda){F}_n(x)$, where $\lambda \in [0, 1]$ and trace the curve, called the \textit{homotopy path}, from a starting point $(x,1)$ to solution point $(x,0)$. Thus, gradually deforming solutions of a starting system $G_n(x) = \mathit{\Theta}_{n}$ into the solutions to the target system ${F}_n(x) = \mathit{\Theta}_{n}$. At each step of the process, a solution of the current system $H_n(x, \tilde{\lambda}) = {\mathit \Theta}_{n}$ is used as a starting solution to the next system $H_n(x, \tilde{\lambda} + \Delta \lambda) = {\mathit \Theta}_{n}$, which is solved using Newton-type methods that require an invertible Jacobian.  In the case of polynomial systems it is trivial to find solutions of the starting system $G_n(x) = \mathit{\Theta}_{n}$. Common approaches for defining $G_n(x)$ include fixed point homotopy, where $G_n(x) = x - x^0$ (where $x^0$ is a starting solution) and Newton homotopy, where $G_n(x) = F_n(x) - F_n(x^0)$. The latter is also referred to a \textit{global homotopy}, where $H_n(x, \lambda) = {F}_n(x) - \lambda {F}_n(x^0)$, where $x^0$ is a starting solution \cite{Allgower2003}. 
An adaptive method for selecting the steps of the homotopy path is presented in \cite{rheinboldt1975}. As per \cite{chen2015homotopy}, advantages of homotopy methods are 1) handling of singular solutions, 2) possibility to obtain multiple solutions using one homotopy path, and 3) preserving Morse indices for gradient systems. 
Paper \cite{Lafmejani20151} introduces a collision-based homotopy continuation technique. Article \cite{chen2012spherical} discusses the problem of divergent homotopy paths and proposes an algorithm which performs projective path tracking. 


Article \cite{10.1145/317275.317286} describes the software package PHCpack for solving polynomial systems using the homotopy method. Other software packages implementing homotopy continuation methods include Bertini \cite{10.5555/2568129} and HomLab~\cite{Numerical_Solution_2005}. Paper \cite{Duff2018} describes a monodromy-based solver.



\section{Interval methods}
\textit{Interval arithmetic} was introduced in article \cite{moore1966interval}. Books on \textit{interval methods} include \cite{neumaier_1991} and \cite{10.5555/1508122}. A real interval $X$ is defined as a set of real numbers between lower and upper bounds 
\[
X = [a,b] = \{x \in \mathbb{R} \,\,|\,\, a \leqslant x \leqslant b\}.
\]
Interval arithmetic is a set of operations such as addition, subtraction, multiplication, and division, defined on the intervals. Further, interval functions, domain and range of which are the intervals, have been defined. In addition to that, interval differentiation and integration have also been proposed. The most important benefit of interval analysis is its accountability for rounding errors due to limited machine precision. If a value is represented as a single number, rounding errors occurring during the computations may accumulate, thus leading to a wrong result. In interval arithmetic a value is represented by a lower and upper bound which provide reliable results during the computations. Further, interval arithmetic can be used to model uncertainty, often arising in practical problems; for example, uncertainty due to imprecise measurements \cite{Kreinovich2016}. Interval arithmetic was standardized by the IEEE in 2015 \cite{IEEE_Interval}.
Global optimization problems can be solved using the \textit{interval branch-and-bound method} which iteratively splits the search space, and removes its parts that do not contain a global solution; multiple splitting schemes have been proposed in the literature \cite{DBLP:reference/opt/RatschekR09}. In this case, application of interval arithmetic allows to guarantee, if a solution exists within a region of interest. Paper \cite{Stuber_etal} proposes to use interval methods for inclusion and exclusion tests, where inclusion tests check existence of the solution within an interval, and exclusion tests check its non-existence.
The \textit{interval Newton method} \cite{Walster2007GlobalOU} is a generalization of a Newton method for interval arithmetic; it can be used to find zeros of a function. Interval Newton method attempts to iteratively narrow down the new interval solution $[x^{k+1}]$ from $[x^{k}]$ via $$[x^{k+1}] = [x^k] \cap \bigl([x^k] - J_{{f}_{n}}\bigl([x^k]\bigr)^{-1}\,{f}_n\bigl([x^k]\bigr)\bigr), \kern0.5cm k=0, 1, 2, \ldots $$ 
Paper \cite{10.2307/2952005} presents an \textit{interval branch-and-prune algorithm} that is capable of finding all solutions of a polynomial system, and the article evaluates the proposed method on large systems with up to 320 variables. 
An interval method was also effectively applied to solve large polynomial systems with up to 2500 variables in \cite{Yamamura2007AnEA}. Article \cite{DBLP:journals/rc/Kubica13} provides a modification of the interval branch-and-bound method and applies it to solving systems of equations. A method of solving nonlinear equations by using interval arithmetic was patented \cite{intervalpatent}. Article \cite{10.1007/978-3-540-39924-7_74} presents details of implementation of a parallel interval optimization algorithm and its application for solving systems of nonlinear equations. A software package for solving equations using interval methods is presented in \cite{JERRELL1998327} and \cite{jerrell2002finding}. Article \cite{eftekhari2020interval} provides interval extensions to Halley's method (discussed in Section \ref{halleyMethod} of this paper), and applies it to finding roots of a single nonlinear equation. Interval methods are utilized to find all solutions of the kinematics SNEs in 
\cite{didrit1998guaranteed,castellet1998algorithm}. Articles \cite{Yamamura1998, YAMAMURA2003587} suggests to use linear programming (LP) for finding all solutions of the SNE; the method in the paper is best suited to SNEs consisting of linear equations with relatively few nonlinear terms. The method is based on the surrounding nonlinear equations by rectangles, based on interval arithmetic. Then, LP can be used to remove parts of the solution space that do not contain the solutions. Paper \cite{YAMAMURA2009405} proposes to extend this method and utilize LP narrowing, that is capable to solve large scale ($n$ = 50 000) separable SNEs. 
Article \cite{kolev1999improved} proposes an extension of the interval method for a SNE based on its transformation to separable form. Paper \cite{Semenov_AL} describes the UniCalc SNE solver software which is based on interval methods. Another software package for interval optimization, that can also be used to solve SNEs is RealPlayer \cite{10.1145/1132973.1132980}.

\section{Synopsis and concluding remarks}
This article presents part one of a survey on methods for solving a system of nonlinear equations (SNE). In part one we have presented a comprehensive survey of methods which can be utilized to search for solutions to a SNE without transforming a SNE into an optimization problem. Since many of the SNEs that arise in real world applications are considered over a finite bounded domain $\mathcal{D}$, 
we first introduced a technique which can be utilized to determine the number of solutions to a SNE that exist within $\mathcal{D}$. Then, we introduced a diverse set of root-finding methods which can be used to search for solutions to a SNE. Next, we described additional methods which have been used to search for solutions to SNEs including methods from symbolic computation, homotopy / continuation methods, and interval methods.




Analyzing this literature has led us to conclude that although there are a variety of root-finding methods which are guaranteed to converge to a solution when applied to SNEs that satisfy certain conditions (such as monotonicity (Eq. \ref{monotonicity}) and Lipschitz continuity (Eq. \ref{l-continuity})), for general SNEs, there is no guarantee that any of the methods described in this paper will converge to a solution in finite time. However, a new taxonomy of SNEs is needed to facilitate the identification of new classes of tractable problems, and to compare the performance of the methods that are most capable of solving them.
We are actively working on the development of such a taxonomy. Additionally, we are actively exploring methods capable of solving systems of nonlinear equations and inequalities. 

In part one of this survey, we discussed methods for solving SNEs without transforming them into optimization problems. In part two of this survey, we will describe various transformations which can be utilized to transform a SNE into an optimization problem, and we will discuss optimization algorithms which can then be used to search for solutions. In part three of this survey, we will present a robust quantitative comparative analysis of methods capable of searching for solutions to SNEs.

\bibliographystyle{unsrt}
\bibliography{LiteratureReview_no_template, symbolic_computation}

\begin{thebibliography}{100}

\bibitem{FLOUDAS2000125}
Christodoulos~A. Floudas.
\newblock Global optimization in design and control of chemical process
  systems.
\newblock {\em Journal of Process Control}, 10(2):125--134, 2000.

\bibitem{SACCO20115424}
Wagner~F. Sacco and N{\'e}lio Henderson.
\newblock Finding all solutions of nonlinear systems using a hybrid
  metaheuristic with fuzzy clustering means.
\newblock {\em Applied Soft Computing}, 11(8):5424--5432, 2011.

\bibitem{Holstad1999}
Astrid Holstad.
\newblock Numerical solution of nonlinear equations in chemical speciation
  calculations.
\newblock {\em Computational Geosciences}, 3(3):229--257, Dec 1999.

\bibitem{Jimenez-Islas2013}
Hugo Jim{\'e}nez-Islas, Gloria~M. Mart{\'i}nez-Gonz{\'a}lez, Jos{\'e}~L.
  Navarrete-Bola{\~{n}}os, Jos{\'e}~E. Botello-{\'A}lvarez, and J.~Manuel
  Oliveros-Mu{\~{n}}oz.
\newblock Nonlinear homotopic continuation methods: A chemical engineering
  perspective review.
\newblock {\em Industrial {\&} Engineering Chemistry Research},
  52(42):14729--14742, Oct 2013.

\bibitem{HENDERSON2010551}
Nélio Henderson, Wagner~F. Sacco, and Gustavo~Mendes Platt.
\newblock Finding more than one root of nonlinear equations via a polarization
  technique: An application to double retrograde vaporization.
\newblock {\em Chemical Engineering Research and Design}, 88(5):551--561, 2010.

\bibitem{Chiang2014}
Hsiao-Dong Chiang, Tian-Qi Zhao, Jiao-Jiao Deng, and Kaoru Koyanagi.
\newblock Homotopy-enhanced power flow methods for general distribution
  networks with distributed generators.
\newblock {\em IEEE Transactions on Power Systems}, 29(1):93--100, 2014.

\bibitem{mehta2016numerical}
Dhagash Mehta, Hung~Dinh Nguyen, and Konstantin Turitsyn.
\newblock Numerical polynomial homotopy continuation method to locate all the
  power flow solutions.
\newblock {\em IET Generation, Transmission \& Distribution},
  10(12):2972--2980, 2016.

\bibitem{Chiang2018}
Hsiao-Dong Chiang and Tao Wang.
\newblock Novel homotopy theory for nonlinear networks and systems and its
  applications to electrical grids.
\newblock {\em IEEE Transactions on Control of Network Systems},
  5(3):1051--1060, 2018.

\bibitem{yang2010fully}
Chao Yang, Jianwen Cao, and Xiao-Chuan Cai.
\newblock A fully implicit domain decomposition algorithm for shallow water
  equations on the cubed-sphere.
\newblock {\em SIAM Journal on Scientific Computing}, 32(1):418--438, 2010.

\bibitem{Schneider2019}
Matti Schneider, Daniel Wicht, and Thomas B{\"o}hlke.
\newblock On polarization-based schemes for the fft-based computational
  homogenization of inelastic materials.
\newblock {\em Computational Mechanics}, 64(4):1073--1095, Oct 2019.

\bibitem{ZHANG2006}
Yunong Zhang.
\newblock A set of nonlinear equations and inequalities arising in robotics and
  its online solution via a primal neural network.
\newblock {\em Neurocomputing}, 70(1):513--524, 2006.

\bibitem{Lafmejani20151}
Amir Salimi~Lafmejani, Ahmad Kalhor, and Mehdi Masouleh.
\newblock A new development of homotopy continuation method, applied in solving
  nonlinear kinematic system of equations of parallel mechanisms.
\newblock {\em 2015 3rd RSI International Conference on Robotics and
  Mechatronics (ICROM)}, 10 2015.

\bibitem{cox2015robotics}
David~A Cox, John Little, and Donal O’Shea.
\newblock Robotics and automatic geometric theorem proving.
\newblock In {\em Ideals, Varieties, and Algorithms}, pages 291--343. Springer,
  2015.

\bibitem{Ji2003KinematicsAO}
Ping Ji and Hongtao Wu.
\newblock Kinematics analysis of an offset 3-upu translational parallel robotic
  manipulator.
\newblock {\em Robotics Auton. Syst.}, 42:117--123, 2003.

\bibitem{cht2016modification}
Strachimir~Cht. Mavrodiev and Maksym Deliyergiyev.
\newblock Modification of the nuclear landscape in the inverse problem
  framework using the generalized bethe-weizs{\"a}cker mass formula.
\newblock {\em arXiv e-prints}, pages arXiv--1602, 2016.

\bibitem{aji2020modified}
Sani Aji, Poom Kumam, Punnarai Siricharoen, Auwal~Bala Abubakar, and
  Mahmoud~Muhammad Yahaya.
\newblock A modified conjugate descent projection method for monotone nonlinear
  equations and image restoration.
\newblock {\em IEEE Access}, 8:158656--158665, 2020.

\bibitem{verschelde1994}
Jan Verschelde, Pierre Verlinden, and Ronald Cools.
\newblock Homotopies exploiting {N}ewton polytopes for solving sparse
  polynomial systems.
\newblock {\em SIAM Journal on Numerical Analysis}, 31(3):915--930, 1994.

\bibitem{Grosan2008}
Crina Grosan and Ajith Abraham.
\newblock A new approach for solving nonlinear equations systems.
\newblock {\em IEEE Transactions on Systems, Man, and Cybernetics - Part A:
  Systems and Humans}, 38(3):698--714, 2008.

\bibitem{golbabai2012}
Ahmad Golbabai, Davood Ahmadian, and Mariyan Milev.
\newblock Radial basis functions with application to finance: American put
  option under jump diffusion.
\newblock {\em Mathematical and Computer Modelling}, 55(3):1354--1362, 2012.

\bibitem{zhang2009existence}
Guang Zhang and Liang Bai.
\newblock Existence of solutions for a nonlinear algebraic system.
\newblock {\em Discrete dynamics in nature and society}, 2009, 2009.

\bibitem{PhysRevLett.81.1195}
Karol Kowalski and Karol Jankowski.
\newblock Towards complete solutions to systems of nonlinear equations of
  many-electron theories.
\newblock {\em Phys. Rev. Lett.}, 81:1195--1198, Aug 1998.

\bibitem{Mehta2011}
Dhagash Mehta.
\newblock Numerical polynomial homotopy continuation method and string vacua.
\newblock {\em Advances in High Energy Physics}, 2011:263937, Oct 2011.

\bibitem{song2020nonlinear}
Yang Song, Chenlin Meng, Renjie Liao, and Stefano Ermon.
\newblock Nonlinear equation solving: A faster alternative to feedforward
  computation.
\newblock {\em arXiv preprint arXiv:2002.03629}, 2020.

\bibitem{pmlr-v108-cai20b}
Yunfeng Cai and Ping Li.
\newblock Solving the robust matrix completion problem via a system of
  nonlinear equations.
\newblock In Silvia Chiappa and Roberto Calandra, editors, {\em Proceedings of
  the Twenty Third International Conference on Artificial Intelligence and
  Statistics}, volume 108 of {\em Proceedings of Machine Learning Research},
  pages 4162--4172. PMLR, 26--28 Aug 2020.

\bibitem{Gatilov2011PropertiesON}
Stepan~Yu. Gatilov.
\newblock Properties of nonlinear systems and convergence of the
  {N}ewton-{R}aphson method in geometric constraint solving.
\newblock volume~32, pages 57–--75. NCC, 2011.

\bibitem{palancz2008dixon}
B{\'e}la Pal{\'a}ncz, Piroska Zaletnyik, Joseph~L Awange, and Erik~W Grafarend.
\newblock Dixon resultant’s solution of systems of geodetic polynomial
  equations.
\newblock {\em Journal of Geodesy}, 82(8):505--511, 2008.

\bibitem{Palancz2009}
B{\'e}la Pal{\'a}ncz, Joseph~L. Awange, Piroska Zaletnyik, and Robert~H. Lewis.
\newblock Linear homotopy solution of nonlinear systems of equations in
  geodesy.
\newblock {\em Journal of Geodesy}, 84(1):79, Sep 2009.

\bibitem{Jansson1998}
Christian Jansson.
\newblock An np-hardness result for nonlinear systems.
\newblock {\em Reliable Computing}, 4(4):345--350, Nov 1998.

\bibitem{Matiyasevich}
Yuri~V. Matiyasevich.
\newblock {\em Hilbert's Tenth Problem}.
\newblock MIT Press, Cambridge, MA, USA, 1993.

\bibitem{khirallah2013solving}
Mustafa~Q Khirallah and MA~Hafiz.
\newblock Solving system of nonlinear equations using family of {J}arratt
  methods.
\newblock {\em International Journal of Differential Equations and
  Applications}, 12(2), 2013.

\bibitem{khirallah2012novel}
MQ~Khirallah and MA~Hafiz.
\newblock Novel three order methods for solving a system of nonlinear
  equations.
\newblock {\em Bulletin of Mathematical Sciences \& Applications}, 1(2):01--14,
  2012.

\bibitem{koupaei2015}
J~Alikhani Koupaei and Seyed Mohammad~Mehdi Hosseini.
\newblock A new hybrid algorithm based on chaotic maps for solving systems of
  nonlinear equations.
\newblock {\em Chaos, Solitons \& Fractals}, 81:233--245, 2015.

\bibitem{el-shorbagy2020}
M.~A. El-Shorbagy and Adel~M. El-Refaey.
\newblock Hybridization of grasshopper optimization algorithm with genetic
  algorithm for solving system of non-linear equations.
\newblock {\em IEEE Access}, 8:220944--220961, 2020.

\bibitem{naidu2016solving}
Y~Ramu Naidu and Akshay~Kumar Ojha.
\newblock Solving multiobjective optimization problems using hybrid cooperative
  invasive weed optimization with multiple populations.
\newblock {\em IEEE Transactions on Systems, Man, and Cybernetics: Systems},
  48(6):821--832, 2016.

\bibitem{kuri2003}
Angel Kuri.
\newblock Solution of simultaneous non-linear equations using genetic
  algorithms.
\newblock {\em WSEAS Transactions on Systems}, 2, 01 2003.

\bibitem{xiao2018solving}
Lin Xiao, Zhijun Zhang, and Shuai Li.
\newblock Solving time-varying system of nonlinear equations by finite-time
  recurrent neural networks with application to motion tracking of robot
  manipulators.
\newblock {\em IEEE Transactions on Systems, Man, and Cybernetics: Systems},
  49(11):2210--2220, 2018.

\bibitem{Geng2009}
Huan-Tong Geng, Yi-Jie Sun, Qing-Xi Song, and Ting-Ting Wu.
\newblock Research of ranking method in evolution strategy for solving
  nonlinear system of equations.
\newblock In {\em 2009 First International Conference on Information Science
  and Engineering}, pages 348--351, 2009.

\bibitem{pei2019}
Jun Pei, Zorica Dra{\v{z}}i{\'c}, Milan Dra{\v{z}}i{\'c}, Nenad Mladenovi{\'c},
  and Panos~M Pardalos.
\newblock Continuous variable neighborhood search (c-vns) for solving systems
  of nonlinear equations.
\newblock {\em INFORMS Journal on Computing}, 31(2):235--250, 2019.

\bibitem{Mousa2008}
Abd~allah Mousa and Islam Eldesoky.
\newblock Genls: Co-evolutionary algorithm for nonlinear system of equations.
\newblock {\em Applied Mathematics and Computation}, 197:633--642, 04 2008.

\bibitem{Gong2017}
Wenyin Gong, Yong Wang, Zhihua Cai, and Shengxiang Yang.
\newblock A weighted biobjective transformation technique for locating multiple
  optimal solutions of nonlinear equation systems.
\newblock {\em IEEE Transactions on Evolutionary Computation}, 21(5):697--713,
  2017.

\bibitem{6849952}
Wu~Song, Yong Wang, Han-Xiong Li, and Zixing Cai.
\newblock Locating multiple optimal solutions of nonlinear equation systems
  based on multiobjective optimization.
\newblock {\em IEEE Transactions on Evolutionary Computation}, 19(3):414--431,
  2015.

\bibitem{Qin2015}
Sha Qin, Sanyou Zeng, Wei Dong, and Xi~Li.
\newblock Nonlinear equation systems solved by many-objective hype.
\newblock In {\em 2015 IEEE Congress on Evolutionary Computation (CEC)}, pages
  2691--2696, 2015.

\bibitem{Guo2020b}
Jinjin Guo, Binbin Qiu, and Yunong Zhang.
\newblock New-type dtz model for solving discrete time-dependent nonlinear
  equation system with robotic-arm application.
\newblock In {\em 2020 10th International Conference on Information Science and
  Technology (ICIST)}, pages 153--162, 2020.

\bibitem{LIAO2020113261}
Zuowen Liao, Wenyin Gong, and Ling Wang.
\newblock Memetic niching-based evolutionary algorithms for solving nonlinear
  equation system.
\newblock {\em Expert Systems with Applications}, 149:113261, 2020.

\bibitem{Liao2020}
Zuowen Liao, Wenyin Gong, Xuesong Yan, Ling Wang, and Chengyu Hu.
\newblock Solving nonlinear equations system with dynamic repulsion-based
  evolutionary algorithms.
\newblock {\em IEEE Transactions on Systems, Man, and Cybernetics: Systems},
  50(4):1590--1601, 2020.

\bibitem{Gao2020}
Weifeng Gao, Yuting Luo, Jingwei Xu, and Shengqi Zhu.
\newblock Evolutionary algorithm with multiobjective optimization technique for
  solving nonlinear equation systems.
\newblock {\em Information Sciences}, 541, 07 2020.

\bibitem{Gong2020c}
Wenyin Gong, Yong Wang, Zhihua Cai, and Ling Wang.
\newblock Finding multiple roots of nonlinear equation systems via a
  repulsion-based adaptive differential evolution.
\newblock {\em IEEE Transactions on Systems, Man, and Cybernetics: Systems},
  50(4):1499--1513, 2020.

\bibitem{Gao2021}
Weifeng Gao, Genghui Li, Qingfu Zhang, Yuting Luo, and Zhenkun Wang.
\newblock Solving nonlinear equation systems by a two-phase evolutionary
  algorithm.
\newblock {\em IEEE Transactions on Systems, Man, and Cybernetics: Systems},
  51(9):5652--5663, 2021.

\bibitem{HE2019104796}
Wei He, Wenyin Gong, Ling Wang, Xuesong Yan, and Chengyu Hu.
\newblock Fuzzy neighborhood-based differential evolution with orientation for
  nonlinear equation systems.
\newblock {\em Knowledge-Based Systems}, 182:104796, 2019.

\bibitem{WU2021106733}
Jianye Wu, Wenyin Gong, and Ling Wang.
\newblock A clustering-based differential evolution with different crowding
  factors for nonlinear equations system.
\newblock {\em Applied Soft Computing}, 98:106733, 2021.

\bibitem{Gao2021b}
Weifeng Gao and Yu~Li.
\newblock Solving a new test set of nonlinear equation systems by evolutionary
  algorithm.
\newblock {\em IEEE Transactions on Cybernetics}, pages 1--10, 2021.

\bibitem{Song2020}
Aijuan Song, Guohua Wu, and Witold Pedrycz.
\newblock Integrating variable reduction strategy with evolutionary algorithm
  for solving nonlinear equations systems.
\newblock {\em CoRR}, abs/2008.04223, 2020.

\bibitem{LIAO2020105312}
Zuowen Liao, Wenyin Gong, Ling Wang, Xuesong Yan, and Chengyu Hu.
\newblock A decomposition-based differential evolution with reinitialization
  for nonlinear equations systems.
\newblock {\em Knowledge-Based Systems}, 191:105312, 2020.

\bibitem{Cheng2018}
Shi Cheng, Junfeng Chen, Xiujuan Lei, and Yuhui Shi.
\newblock Locating multiple optima via brain storm optimization algorithms.
\newblock {\em IEEE Access}, 6:17039--17049, 2018.

\bibitem{gong2021nonlinear}
Wenyin Gong, Zuowen Liao, Xianyan Mi, Ling Wang, and Yuanyuan Guo.
\newblock Nonlinear equations solving with intelligent optimization algorithms:
  A survey.
\newblock {\em Complex System Modeling and Simulation}, 1(1):15--32, 2021.

\bibitem{Ahookhosh2013}
Masoud Ahookhosh, Keyvan Amini, and Somayeh Bahrami.
\newblock Two derivative-free projection approaches for systems of large-scale
  nonlinear monotone equations.
\newblock {\em Numerical Algorithms}, 64, 10 2013.

\bibitem{TanabeV2006}
Susumu Tanab{\'e} and Michael~N. Vrahatis.
\newblock On perturbation of roots of homogeneous algebraic systems.
\newblock {\em Mathematics of Computation}, 255:1383--1402, 2006.

\bibitem{Ullah2021}
Najib Ullah, Jamilu Sabi'u, and Abdullah Shah.
\newblock A derivative-free scaling memoryless broyden–
  fletcher–goldfarb–shanno method for solving a system of monotone
  nonlinear equations.
\newblock {\em Numerical Linear Algebra with Applications}, 28(5):e2374, 2021.

\bibitem{aji2021}
Sani Aji, Poom Kumam, Aliyu~Muhammed Awwal, Mahmoud~Muhammad Yahaya, and Wiyada
  Kumam.
\newblock Two hybrid spectral methods with inertial effect for solving system
  of nonlinear monotone equations with application in robotics.
\newblock {\em IEEE Access}, 9:30918--30928, 2021.

\bibitem{Martinez1994}
Jos{\'e}~Mario Mart{\'i}nez.
\newblock {\em Algorithms for Solving Nonlinear Systems of Equations}, pages
  81--108.
\newblock Springer Netherlands, Dordrecht, 1994.

\bibitem{OrtegaR2000}
James~M. Ortega and Werner~C. Rheinboldt.
\newblock {\em Iterative Solution of Nonlinear Equations in Several Variables}.
\newblock Society for Industrial and Applied Mathematics, January 2000.

\bibitem{Picard1892}
{\'E}mile Picard.
\newblock Sur le nombre des racines communes \`a plusieurs \'equations
  simultan\'ees.
\newblock {\em Journal de Math\'ematiques Pures et Appliqu\'ees}, 8($4^e$
  s\'erie):5--24, 1892.

\bibitem{Picard1922}
{\'E}mile Picard.
\newblock {\em Trait\'e d'Analyse}.
\newblock Gauthier-Villars, Paris, 1922.

\bibitem{Cronin1964}
Jane Cronin.
\newblock {\em Fixed Points and Topological Degree in Nonlinear Analysis}.
\newblock American Mathematical Society, Providence, Rhode Island, 1964.

\bibitem{Lloyd1978}
Noel~G. Lloyd.
\newblock {\em Degree Theory}.
\newblock Oxford University Press, New York, 1978.

\bibitem{OReganCC2006}
Donal O'Regan, Yeol~Je Cho, and Yu-Qing Chen.
\newblock {\em Topological Degree Theory and Applications}.
\newblock Taylor \& Francis Group, Boca Raton, FL, 2006.

\bibitem{OutereloR2009}
Enrique Outerelo and Jes{\'u}s~M. Ruiz.
\newblock {\em Mapping Degree Theory}.
\newblock American Mathematical Society, Providence, Rhode Island, 2009.

\bibitem{Sikorski2001}
Krzysztof~A. Sikorski.
\newblock {\em Optimal Solution of Nonlinear Equations}.
\newblock Oxford University Press, New York, 2001.

\bibitem{BergaminBV2002}
Jeroen~M. Bergamin, Tassos~C. Bountis, and Michael~N. Vrahatis.
\newblock Homoclinic orbits of invertible maps.
\newblock {\em Nonlinearity}, 15(5):1603--1619, 2002.

\bibitem{EmirisMV1999}
Ioannis~Z. Emiris, Bernard Mourrain, and Michael~N. Vrahatis.
\newblock Sign methods for counting and computing real roots of algebraic
  systems.
\newblock Technical report, Rapport de Recherche No.3669, INRIA (Institut
  National de Recherche en Informatique et en Automatique), Sophia Antipolis,
  France, April 1999.

\bibitem{KavvadiasV1996}
Dimitris~J. Kavvadias and Michael~N. Vrahatis.
\newblock Locating and computing all the simple roots and extrema of a
  function.
\newblock {\em SIAM Journal on Scientific Computing}, 17(5):1232--1248, 19965.

\bibitem{Kearfott1979}
R.~Baker Kearfott.
\newblock An efficient degree-computation method for a generalized method of
  bisection.
\newblock {\em Numerische Mathematik}, 32:109--127, 1979.

\bibitem{MourrainPTV2006}
Bernard Mourrain, Nicos~G. Pavlidis, Dimitris~K. Tasoulis, and Michael~N.
  Vrahatis.
\newblock Determining the number of real roots of polynomials through neural
  networks.
\newblock {\em Computers and Mathematics with Applications}, 51(3-4):527--536,
  2006.

\bibitem{MourrainVY2002}
Bernard Mourrain, Michael~N. Vrahatis, and Yakoubsohn Jean-Claude.
\newblock On the complexity of isolating real roots and computing with
  certainty the topological degree.
\newblock {\em Journal of Complexity}, 18(2):612--640, 2002.

\bibitem{PlagianakosNV2001}
Vassilis~P. Plagianakos, Nicos~K. Nousis, and Michael~N. Vrahatis.
\newblock Locating and computing in parallel all the simple roots of special
  functions using pvm.
\newblock {\em Journal of Computational and Applied Mathematics},
  133(1-2):545--554, 2001.

\bibitem{PolymilisSSTV2003}
Chronis Polymilis, Graziano Servizi, Charalampos Skokos, Giorgio Turchetti, and
  Michael~N. Vrahatis.
\newblock Topological degree theory and local analysis of area preserving maps.
\newblock {\em Chaos}, 13(1):94--104, 2003.

\bibitem{Stenger1975}
Frank Stenger.
\newblock Computing the topological degree of a mapping in ${R}^n$.
\newblock {\em Numerische Mathematik}, 25:23--38, 1975.

\bibitem{Stynes1979}
Martin Stynes.
\newblock An algorithm for numerical calculation of topological degree.
\newblock {\em Applicable Analysis}, 25:23--38, 1979.

\bibitem{Vrahatis1995}
Michael~N. Vrahatis.
\newblock An efficient method for locating and computing periodic orbits of
  nonlinear mappings.
\newblock {\em Journal of Computational Physics}, 119:105--119, 1995.

\bibitem{Vrahatis1988a}
Michael~N. Vrahatis.
\newblock Solving systems of nonlinear equations using the nonzero value of the
  topological degree.
\newblock {\em ACM Transactions on Mathematical Software}, 14(4):312--329, Dec.
  1988.

\bibitem{Vrahatis1989}
Michael~N. Vrahatis.
\newblock A short proof and a generalization of {M}iranda's existence theorem.
\newblock {\em Proceedings of the American Mathematical Society},
  107(3):701--703, 1989.

\bibitem{VrahatisI1986}
Michael~N. Vrahatis and Kosmas~I. Iordanidis.
\newblock A rapid generalized method of bisection for solving systems of
  nonlinear equations.
\newblock {\em Numerische Mathematik}, 49(2-3):123--138, 1986.

\bibitem{VrahatisSTB1993}
Michael~N. Vrahatis, Graziano Servizi, Giorgio Turchetti, and Tassos~C.
  Bountis.
\newblock A procedure to compute the fixed points and visualize the orbits of a
  2{D} map.
\newblock Technical report, CERN SL/93-06 (AP), CERN-SL Division, European
  Organization for Nuclear Research (CERN), Geneva, Switzerland, February 1993.

\bibitem{Franek2015}
Peter Franek and Marek Kr\v{c}\'{a}l.
\newblock Robust satisfiability of systems of equations.
\newblock {\em J. ACM}, 62(4), sep 2015.

\bibitem{eustaquio2018new}
Rodrigo~G Eustaquio, Ademir~A Ribeiro, and Miguel~A Dumett.
\newblock A new class of root-finding methods in $r^n$: the inexact tensor-free
  chebyshev--halley class.
\newblock {\em Computational and Applied Mathematics}, 37(5):6654--6675, 2018.

\bibitem{DennisS1996}
John~E. Dennis and Robert~B. Schnabel.
\newblock {\em Numerical methods for unconstrained optimization and nonlinear
  equations}.
\newblock SIAM, 1996.

\bibitem{kelley2003}
Carl~T. Kelley.
\newblock {\em Solving nonlinear equations with Newton's method}.
\newblock SIAM, 2003.

\bibitem{rheinboldt1998}
Werner~C. Rheinboldt.
\newblock {\em Methods for solving systems of nonlinear equations}.
\newblock SIAM, 1998.

\bibitem{frontini2004}
Marco Frontini and E~Sormani.
\newblock Third-order methods from quadrature formulae for solving systems of
  nonlinear equations.
\newblock {\em Applied Mathematics and Computation}, 149(3):771--782, 2004.

\bibitem{darvishi2007b}
MT~Darvishi and Ali Barati.
\newblock A fourth-order method from quadrature formulae to solve systems of
  nonlinear equations.
\newblock {\em Applied Mathematics and Computation}, 188(1):257--261, 2007.

\bibitem{srivastava2021}
Hari~M Srivastava, Javed Iqbal, Muhammad Arif, Alamgir Khan, Yusif~S Gasimov,
  and Ronnason Chinram.
\newblock A new application of gauss quadrature method for solving systems of
  nonlinear equations.
\newblock {\em Symmetry}, 13(3):432, 2021.

\bibitem{ASLAMNOOR2009}
Muhammad {Aslam Noor} and Muhammad Waseem.
\newblock Some iterative methods for solving a system of nonlinear equations.
\newblock {\em Computers \& Mathematics with Applications}, 57(1):101--106,
  2009.

\bibitem{Ben-Israel1965}
Adi Ben-Israel.
\newblock A modified newton-raphson method for the solution of systems of
  equations.
\newblock {\em Israel Journal of Mathematics}, 3(2):94--98, Jun 1965.

\bibitem{Yudin2021}
N.~E. Yudin.
\newblock Adaptive gauss--newton method for solving systems of nonlinear
  equations.
\newblock {\em Doklady Mathematics}, 104(2):293--296, Sep 2021.

\bibitem{BENISRAEL1966243}
Adi Ben-Israel.
\newblock A newton-raphson method for the solution of systems of equations.
\newblock {\em Journal of Mathematical Analysis and Applications},
  15(2):243--252, 1966.

\bibitem{HUESO200977}
José~L. Hueso, Eulalia Martínez, and Juan~R. Torregrosa.
\newblock Modified newton’s method for systems of nonlinear equations with
  singular jacobian.
\newblock {\em Journal of Computational and Applied Mathematics},
  224(1):77--83, 2009.

\bibitem{Gatilov2014}
Stepan~Yu. Gatilov.
\newblock Using low-rank approximation of the {J}acobian matrix in the
  {N}ewton-{R}aphson method to solve certain singular equations.
\newblock {\em J. Comput. Appl. Math.}, 272:8--24, 2014.

\bibitem{xu2007a}
Xiubin Xu and Chong Li.
\newblock Convergence of newton's method for systems of equations with constant
  rank derivatives.
\newblock {\em Journal of Computational Mathematics}, 25(6):705--718, 2007.

\bibitem{grapsa1989}
Theodoula~N. Grapsa and Michael~N. Vrahatis.
\newblock The implicit function theorem for solving systems of nonlinear
  equations in ${R}^2$.
\newblock {\em International Journal of Computer Mathematics},
  28(1-4):171--181, 1989.

\bibitem{grapsa1990}
Theodoula~N. Grapsa and Michael~N. Vrahatis.
\newblock A dimension-reducing method for solving systems of nonlinear
  equations in ${R}^n$.
\newblock {\em International Journal of Computer Mathematics},
  32(3-4):205--216, 1990.

\bibitem{grapsa1995}
Theodoula~N. Grapsa and Michael~N. Vrahatis.
\newblock A new dimension-reducing method for solving systems of nonlinear
  equations.
\newblock {\em International Journal of Computer Mathematics}, 55:235--244,
  1995.

\bibitem{grapsa1996}
Theodoula~N. Grapsa and Michael~N. Vrahatis.
\newblock A dimension-reducing method for unconstrained optimization.
\newblock {\em Journal of Computational and Applied Mathematics}, 66:239--253,
  1996.

\bibitem{grapsaVB1990}
Theodoula~N. Grapsa, Michael~N. Vrahatis, and Tassos~C. Bountis.
\newblock Solving systems of nonlinear equations in ${R}^n$ using a rotating
  hyperplane in ${R}^{n+1}$.
\newblock {\em International Journal of Computer Mathematics},
  35(1-4):133--151, 1990.

\bibitem{broyden1965}
Charles~G. Broyden.
\newblock A class of methods for solving nonlinear simultaneous equations.
\newblock {\em Mathematics of Computation}, 19(92):577--593, 1965.

\bibitem{GomesRuggiero1992}
M{\'{a}}rcia~A. Gomes-Ruggiero, Jos{\'{e}}~Mario Mart{\'{\i}}nez, and
  Antonio~Carlos Moretti.
\newblock Comparing algorithms for solving sparse nonlinear systems of
  equations.
\newblock {\em {SIAM} Journal on Scientific and Statistical Computing},
  13(2):459--483, March 1992.

\bibitem{epitropakis2011}
Michael~G. Epitropakis and Michael~N. Vrahatis.
\newblock Studying the basin of convergence of methods for computing periodic
  orbits.
\newblock {\em International Journal of Bifurcation and Chaos},
  21(08):2079--2106, 2011.

\bibitem{Saad1986GMRES}
Youcef Saad and Martin~H. Schultz.
\newblock Gmres: A generalized minimal residual algorithm for solving
  nonsymmetric linear systems.
\newblock {\em SIAM Journal on Scientific and Statistical Computing},
  7(3):856--869, 1986.

\bibitem{Arnoldi1951ThePO}
Walter~E. Arnoldi.
\newblock The principle of minimized iterations in the solution of the matrix
  eigenvalue problem.
\newblock {\em Quarterly of Applied Mathematics}, 9:17--29, 1951.

\bibitem{sikorski1982}
Krzysztof~A Sikorski.
\newblock Bisection is optimal.
\newblock {\em Numerische Mathematik}, 40:111--117, 1982.

\bibitem{KavvadiasMV1999}
Dimitris~J. Kavvadias, Frosso~S. Makri, and Michael~N. Vrahatis.
\newblock Locating and computing arbitrarily distributed zeros.
\newblock {\em SIAM Journal on Scientific Computing}, 21(3):954--969, 1999.

\bibitem{KavvadiasMV2005}
Dimitris~J. Kavvadias, Frosso~S. Makri, and Michael~N. Vrahatis.
\newblock Efficiently computing many roots of a function.
\newblock {\em SIAM Journal on Scientific Computing}, 27(1):93--107, 2005.

\bibitem{Vrahatis1986}
Michael~N. Vrahatis.
\newblock An error estimation for the method of bisection in ${R}^n$.
\newblock {\em Bulletin of the Greek Mathematical Society}, 27:161--174, 1986.

\bibitem{Vrahatis1988b}
Michael~N. Vrahatis.
\newblock {CHABIS}: A mathematical software package for locating and evaluating
  roots of systems of nonlinear equations.
\newblock {\em ACM Transactions on Mathematical Software}, 14(4):330--336, Dec.
  1988.

\bibitem{VrahatisAM1996}
Michael~N. Vrahatis, George~S. Androulakis, and George~E. Manoussakis.
\newblock A new unconstrained optimization method for imprecise function and
  gradient values.
\newblock {\em Journal of Mathematical Analysis and Applications},
  197(2):586--607, 1996.

\bibitem{VrahatisIB1997}
Michael~N. Vrahatis, Heinz Isliker, and Tassos~C. Bountis.
\newblock Structure and breakdown of invariant tori in a 4-{D} mapping model of
  accelerator dynamics.
\newblock {\em International Journal of Bifurcation and Chaos},
  7(12):2707--2722, 1997.

\bibitem{ZottouKMV2018}
Dimitra-Nefeli~A. Zottou, Dimitris~J. Kavvadias, Frosso~S. Makri, and
  Michael~N. Vrahatis.
\newblock Manbis--a {C}++ mathematical software package for locating and
  computing efficiently many roots of a function: Theoretical issues.
\newblock {\em ACM Transactions on Mathematical Software}, 44(3):1--7, April
  2018.

\bibitem{AlexandropoulosPV2020}
Stamatios-Aggelos~N. Alexandropoulos, Panos~M. Pardalos, and Michael~N.
  Vrahatis.
\newblock Dynamic search trajectory methods for global optimization.
\newblock {\em Annals of Mathematics and Artificial Intelligence},
  88(1-3):3--37, 2020.

\bibitem{BoutsinasV2001}
Bassilis Boutsinas and Michael~N. Vrahatis.
\newblock Artificial nonmonotonic neural networks.
\newblock {\em Artificial Intelligence}, 132(1):1--38, 2001.

\bibitem{VrahatisALM2000}
Michael~N. Vrahatis, George~S. Androulakis, John~N. Lambrinos, and George~D.
  Magoulas.
\newblock A class of gradient unconstrained minimization algorithms with
  adaptive stepsize.
\newblock {\em Journal of Computational and Applied Mathematics},
  114(2):367--386, 2000.

\bibitem{VrahatisMP2000}
Michael~N. Vrahatis, George~D. Magoulas, and Vassilis~P Plagianakos.
\newblock Globally convergent modification of the quickprop method.
\newblock {\em Neural Processing Letters}, 12(2):159--169, 2000.

\bibitem{HERNANDEZ199927}
M.A. Hernández and M.A. Salanova.
\newblock Indices of convexity and concavity. application to halley method.
\newblock {\em Applied Mathematics and Computation}, 103(1):27--49, 1999.

\bibitem{hernandez1993family}
MA~Hern{\'a}ndez and MA~Salanova.
\newblock A family of chebyshev-halley type methods.
\newblock {\em International Journal of Computer Mathematics}, 47(1-2):59--63,
  1993.

\bibitem{gutierrez_hernandez_1997}
J.M. Gutiérrez and M.A. Hernández.
\newblock A family of chebyshev-halley type methods in banach spaces.
\newblock {\em Bulletin of the Australian Mathematical Society},
  55(1):113–130, 1997.

\bibitem{babajee2010}
DKR Babajee, MZ~Dauhoo, MT~Darvishi, A~Karami, and Ali Barati.
\newblock Analysis of two chebyshev-like third order methods free from second
  derivatives for solving systems of nonlinear equations.
\newblock {\em Journal of Computational and Applied Mathematics},
  233(8):2002--2012, 2010.

\bibitem{STEIHAUG2013230}
Trond Steihaug and Sara Suleiman.
\newblock Rate of convergence of higher order methods.
\newblock {\em Applied Numerical Mathematics}, 67:230--242, 2013.
\newblock NUMAN 2010.

\bibitem{Argyros1994}
Ioannis~K. Argyros, Dong Chen, and Qian Qingshan.
\newblock The jarratt method in banach space setting.
\newblock {\em Journal of Computational and Applied Mathematics},
  51(1):103--106, May 1994.

\bibitem{ahmad2015}
Fayyaz Ahmad, Emran Tohidi, Malik~Zaka Ullah, and Juan~A Carrasco.
\newblock Higher order multi-step jarratt-like method for solving systems of
  nonlinear equations: Application to pdes and odes.
\newblock {\em Computers \& Mathematics with Applications}, 70(4):624--636,
  2015.

\bibitem{zhanlav2021}
Thugal Zhanlav and Khuder Otgondorj.
\newblock Higher order jarratt-like iterations for solving systems of nonlinear
  equations.
\newblock {\em Applied Mathematics and Computation}, 395:125849, 2021.

\bibitem{frank1984tensor}
Paul~David Frank.
\newblock {\em Tensor methods for solving systems of nonlinear equations
  (numerical analysis, optimization)}.
\newblock PhD thesis, University of Colorado at Boulder, 1984.

\bibitem{schnabel1986solving}
Robert~B Schnabel and Paul~D Frank.
\newblock Solving systems of nonlinear equations by tensor methods.
\newblock Technical report, Department of Computer Science, Colorado University
  at Boulder, 1986.

\bibitem{bouaricha1992solving}
Ali Bouaricha.
\newblock {\em Solving large sparse systems of nonlinear equations and
  nonlinear least squares problems using tensor methods on sequential and
  parallel computers}.
\newblock PhD thesis, University of Colorado at Boulder, 1992.

\bibitem{bouaricha1994tensor}
Ali Bouaricha and Robert~B Schnabel.
\newblock Tensor methods for large, sparse systems of nonlinear equations.
\newblock {\em Preprint MCS-P473-1094, Mathematics and Computer Science
  Division, Argonne National Laboratory}, 1994.

\bibitem{feng1997tensor}
Dan Feng and Thomas~H Pulliam.
\newblock Tensor-gmres method for large systems of nonlinear equations.
\newblock {\em SIAM Journal on Optimization}, 7(3):757--779, 1997.

\bibitem{bouaricha1997algorithm}
Ali Bouaricha and Robert~B Schnabel.
\newblock Algorithm 768: {TENSOLVE}: A software package for solving systems of
  nonlinear equations and nonlinear least-squares problems using tensor
  methods.
\newblock {\em ACM Transactions on Mathematical Software (TOMS)},
  23(2):174--195, 1997.

\bibitem{bouaricha1998tensor}
Ali Bouaricha and Robert~B Schnabel.
\newblock Tensor methods for large sparse systems of nonlinear equations.
\newblock {\em Mathematical programming}, 82(3):377--400, 1998.

\bibitem{bader2004performance}
Brett~W Bader and Robert~B Schnabel.
\newblock On the performance of tensor methods for solving ill-conditioned
  problems.
\newblock {\em SIAM Journal on Scientific Computing}, 29(6):2329--2351, 2004.

\bibitem{Bader2005}
Brett~W. Bader.
\newblock Tensor-{K}rylov methods for solving large-scale systems of nonlinear
  equations.
\newblock {\em SIAM Journal on Numerical Analysis}, 43(3):1321--1347, 2005.

\bibitem{Brezinski2004ACO}
Claude Brezinski.
\newblock A classification of quasi-{N}ewton methods.
\newblock {\em Numerical Algorithms}, 33:123--135, 2004.

\bibitem{Li1989}
Guangye Li.
\newblock Successive column correction algorithms for solving sparse nonlinear
  systems of equations.
\newblock {\em Mathematical Programming}, 43(1):187--207, Jan 1989.

\bibitem{Mario1992}
Jos{\'e}~Mario Mart{\'\i}nez and Mário~C. Zambaldi.
\newblock An inverse column-updating method for solving large–scale nonlinear
  systems of equations.
\newblock {\em Optimization Methods and Software}, 1(2):129--140, 1992.

\bibitem{broyden1970}
Charles~G. Broyden.
\newblock {The convergence of a class of double-rank minimization algorithms 1.
  General considerations}.
\newblock {\em IMA Journal of Applied Mathematics}, 6(1):76--90, 03 1970.

\bibitem{fletcher1970}
Roger Fletcher.
\newblock {A new approach to variable metric algorithms}.
\newblock {\em The Computer Journal}, 13(3):317--322, 01 1970.

\bibitem{Goldfarb1970}
Donald Goldfarb.
\newblock A family of variable-metric methods derived by variational means.
\newblock {\em Mathematics of Computation}, 24(109):23--26, 1970.

\bibitem{shanno1970conditioning}
David~F Shanno.
\newblock Conditioning of quasi-newton methods for function minimization.
\newblock {\em Mathematics of computation}, 24(111):647--656, 1970.

\bibitem{Liu1989}
Dong~C. Liu and Jorge Nocedal.
\newblock On the limited memory bfgs method for large scale optimization.
\newblock {\em Mathematical Programming}, 45(1):503--528, Aug 1989.

\bibitem{birgin2003}
Ernesto~G Birgin, Nata{\v{s}}a Kreji{\'c}, and Jos{\'e}~Mario Mart{\'\i}nez.
\newblock Globally convergent inexact quasi-newton methods for solving
  nonlinear systems.
\newblock {\em Numerical algorithms}, 32(2):249--260, 2003.

\bibitem{dauda2020}
MK~Dauda, AS~Magaji, US~Shehub, MA~Usman, and MY~Waziri.
\newblock A simple conjugate gradient type method for solving large-scale
  systems of nonlinear equations.
\newblock {\em Malaysian Journal of Computing and Applied Mathematics},
  3(2):25--35, 2020.

\bibitem{ek2020}
David Ek.
\newblock {\em Approaches to accelerate methods for solving systems of
  equations arising in nonlinear optimization}.
\newblock PhD thesis, KTH Royal Institute of Technology, 2020.

\bibitem{kamfa2020}
K~Kamfa, MY~Waziri, IM~Sulaiman, M~Mamat, and Hery Ibrahim.
\newblock A quasi-newton like method via modified rational approximation model
  for solving system of nonlinear equation.
\newblock {\em Journal of Advanced Research in Dynamical and Control Systems},
  2020.

\bibitem{leong2011}
Wah~June Leong, Malik~Abu Hassan, and Muhammad~Waziri Yusuf.
\newblock A matrix-free quasi-newton method for solving large-scale nonlinear
  systems.
\newblock {\em Computers \& Mathematics with Applications}, 62(5):2354--2363,
  2011.

\bibitem{luo2008}
Ya-Zhong Luo, Guo-Jin Tang, and Li-Ni Zhou.
\newblock Hybrid approach for solving systems of nonlinear equations using
  chaos optimization and quasi-newton method.
\newblock {\em Applied Soft Computing}, 8(2):1068--1073, 2008.

\bibitem{mamat2016}
Mustafa Mamat, MK~Dauda, MY~Waziri, Fadhilah Ahmad, and Fatma~Susilawati
  Mohamad.
\newblock Improved quasi-newton method via psb update for solving systems of
  nonlinear equations.
\newblock In {\em AIP Conference Proceedings}, volume 1782, page 030009. AIP
  Publishing LLC, 2016.

\bibitem{MARTINEZ2000}
José~Mario Mart{\'{\i}}nez.
\newblock Practical quasi-newton methods for solving nonlinear systems.
\newblock {\em Journal of Computational and Applied Mathematics},
  124(1):97--121, 2000.
\newblock Numerical Analysis 2000. Vol. IV: Optimization and Nonlinear
  Equations.

\bibitem{perez2004}
Rosana P{\'e}rez and V{\'e}ra Lucia~Rocha Lopes.
\newblock Recent applications and numerical implementation of quasi-newton
  methods for solving nonlinear systems of equations.
\newblock {\em Numerical Algorithms}, 35(2):261--285, 2004.

\bibitem{papini2020}
Alessandra Papini, Margherita Porcelli, and Cristina Sgattoni.
\newblock On the global convergence of a new spectral residual algorithm for
  nonlinear systems of equations.
\newblock {\em Bollettino dell'Unione Matematica Italiana}, pages 1--12, 2020.

\bibitem{LaCruz2006}
William~La Cruz, Jos{\'{e}}~Mario Mart{\'{\i}}nez, and Marcos Raydan.
\newblock Spectral residual method without gradient information for solving
  large-scale nonlinear systems of equations.
\newblock {\em Mathematics of Computation}, 75(255):1429--1449, April 2006.

\bibitem{bai2003}
Zhong-Zhi Bai, Gene~H. Golub, and Michael~K. Ng.
\newblock Hermitian and skew-hermitian splitting methods for non-hermitian
  positive definite linear systems.
\newblock {\em SIAM Journal on Matrix Analysis and Applications},
  24(3):603--626, 2003.

\bibitem{wu2013}
Qingbiao Wu and Minhong Chen.
\newblock Convergence analysis of modified newton-hss method for solving
  systems of nonlinear equations.
\newblock {\em Numerical Algorithms}, 64(4):659--683, 2013.

\bibitem{bai2009}
Zhong-Zhi Bai and Xi~Yang.
\newblock On hss-based iteration methods for weakly nonlinear systems.
\newblock {\em Applied numerical mathematics}, 59(12):2923--2936, 2009.

\bibitem{amiri2019}
Abdolreza Amiri, Alicia Cordero, Mohammad~Taghi Darvishi, and Juan~R
  Torregrosa.
\newblock A fast algorithm to solve systems of nonlinear equations.
\newblock {\em Journal of Computational and Applied Mathematics}, 354:242--258,
  2019.

\bibitem{Levenberg1944}
Kenneth Levenberg.
\newblock A method for the solution of certain non-linear problems in least
  squares.
\newblock {\em Quarterly of Applied Mathematics}, 2(2):164--168, 1944.

\bibitem{Marquardt1963}
Donald Marquardt.
\newblock An algorithm for least-squares estimation of nonlinear parameters.
\newblock {\em SIAM Journal on Applied Mathematics}, 11(2):431--441, 1963.

\bibitem{Fletcher1971}
Roger Fletcher.
\newblock A modified marquardt subroutine for nonlinear least squares.
\newblock Technical report, Atomic Energy Research Establishment, Harwell,
  England (United Kingdom), 1971.

\bibitem{kuzina2020}
Valentina Kuzina and Alexander Koshev.
\newblock Modification of the levenberg--marquardt algorithm for solving
  complex computational construction problems.
\newblock In {\em IOP Conference Series: Materials Science and Engineering},
  volume 960, page 032039. IOP Publishing, 2020.

\bibitem{Rubio2021}
Jos{\'e} de~Jes{\'u}s Rubio.
\newblock Stability analysis of the modified levenberg–marquardt algorithm
  for the artificial neural network training.
\newblock {\em IEEE Transactions on Neural Networks and Learning Systems},
  32(8):3510--3524, 2021.

\bibitem{BrownG1971}
Kenneth~M. Brown and William~B. Gearhart.
\newblock Deflation techniques for the calculation of further solutions of a
  nonlinear system.
\newblock {\em Numerische Mathematik}, 16, 1971.

\bibitem{Wilkinson1994}
James~H. Wilkinson.
\newblock {\em Rounding errors in algebraic processes}.
\newblock Dover Publications, Inc., New York, 1994.

\bibitem{KalantonisPPRV2003}
Vassilis~S. Kalantonis, Efstathios~A. Perdios, Angela~E. Perdiou, Omiros Ragos,
  and Michael~N. Vrahatis.
\newblock Deflation techniques for the determination of periodic solutions of a
  certain period.
\newblock {\em Astrophysics and Space Science}, 288(2):591--599, 2003.

\bibitem{ParsopoulosPMV2001}
Konstantinos~E. Parsopoulos, Vasssilis~P. Plagianakos, George~D. Magoulas, and
  Michael~N. Vrahatis.
\newblock Objective function ``stretching'' to alleviate convergence to local
  minima.
\newblock {\em Nonlinear Analysis – Theory, Methods \& Applications},
  47(5):3419--3424, 2001.

\bibitem{ParsopoulosV2002}
Konstantinos~E. Parsopoulos and Michael~N. Vrahatis.
\newblock Recent approaches to global optimization problems through particle
  swarm optimization.
\newblock {\em Natural Computing}, 1(2-3):235--306, 2002.

\bibitem{ParsopoulosV2004}
Konstantinos~E. Parsopoulos and Michael~N. Vrahatis.
\newblock On the computation of all global minimizers through particle swarm
  optimization.
\newblock {\em IEEE Transactions on Evolutionary Computation}, 8(3):211--224,
  2004.

\bibitem{ParsopoulosV2010}
Konstantinos~E. Parsopoulos and Michael~N. Vrahatis.
\newblock {\em Particle swarm optimization and intelligence: Advances and
  applications}.
\newblock Information Science Publishing (IGI Global), Hershey, PA, USA, 2010.

\bibitem{ParsopoulosV2021}
Konstantinos~E. Parsopoulos and Michael~N. Vrahatis.
\newblock Deflection and stretching techniques for detection of multiple
  minimizers in multimodal optimization problems.
\newblock In {\em Metaheuristics for Finding Multiple Solutions, M. Preuss,
  M.G. Epitropakis, J.E. Fieldsend and X. Li (eds.), Chapter 6}, pages
  129--144. Natural Computing Series, Springer Nature Switcherland AG, 2021.

\bibitem{GeddesCzaporLabahn:1992}
K.~O. Geddes, S.~R. Czapor, and G.~Labahn.
\newblock {\em Algorithms for computer algebra}.
\newblock Kluwer Academic Publishers, Boston, MA, 1992.

\bibitem{Wester:1999}
Michael~J. Wester.
\newblock {\em Computer Algebra Systems: A Practical Guide}.
\newblock John Wiley \& Sons, Chichester, United Kingdom, 1999.
\newblock \url{https://math.unm.edu/~wester/cas/book/contents.html}.

\bibitem{Davenport:Siret:Tournier:1993}
J.~H. Davenport, Y.~Siret, and E.~Tournier.
\newblock {\em Computer algebra}.
\newblock Academic Press, Ltd., London, second edition, 1993.
\newblock Systems and algorithms for algebraic computation, With a preface by
  Daniel Lazard, Translated from the French by A. Davenport and J. H.
  Davenport, With a foreword by Anthony C. Hearn.

\bibitem{vzGathen:Gerhard:2013}
Joachim von~zur Gathen and J\"{u}rgen Gerhard.
\newblock {\em Modern computer algebra}.
\newblock Cambridge University Press, Cambridge, third edition, 2013.

\bibitem{Teo_Mora:EMA_I_2003}
Teo Mora.
\newblock {\em Solving polynomial equation systems. {I}}, volume~88 of {\em
  Encyclopedia of Mathematics and its Applications}.
\newblock Cambridge University Press, Cambridge, 2003.
\newblock The Kronecker-Duval philosophy.

\bibitem{Teo_Mora:EMA_II_2005}
Teo Mora.
\newblock {\em Solving polynomial equation systems. {II}}, volume~99 of {\em
  Encyclopedia of Mathematics and its Applications}.
\newblock Cambridge University Press, Cambridge, 2005.
\newblock Macaulay's paradigm and Gr\"{o}bner technology.

\bibitem{Teo_Mora:EMA_III_2015}
Teo Mora.
\newblock {\em Solving polynomial equation systems. {V}ol. {III}. {A}lgebraic
  solving}, volume 157 of {\em Encyclopedia of Mathematics and its
  Applications}.
\newblock Cambridge University Press, Cambridge, 2015.

\bibitem{Teo_Mora:EMA_IV_2016}
Teo Mora.
\newblock {\em Solving polynomial equation systems. {V}ol. {IV}. {B}uchberger
  theory and beyond}, volume 158 of {\em Encyclopedia of Mathematics and its
  Applications}.
\newblock Cambridge University Press, Cambridge, 2016.

\bibitem{William_W_Adams_Philippe_Loustaunau:1994}
William~W. Adams and Philippe Loustaunau.
\newblock {\em An introduction to {G}r\"{o}bner bases}, volume~3 of {\em
  Graduate Studies in Mathematics}.
\newblock American Mathematical Society, Providence, RI, 1994.

\bibitem{Thomas_Becker_Volker_Weispfenning:1993}
Thomas Becker and Volker Weispfenning.
\newblock {\em Gr\"{o}bner bases}, volume 141 of {\em Graduate Texts in
  Mathematics}.
\newblock Springer-Verlag, New York, 1993.
\newblock A computational approach to commutative algebra, In cooperation with
  Heinz Kredel.

\bibitem{Ralf_Froberg:1997}
Ralf Fr\"{o}berg.
\newblock {\em An introduction to {G}r\"{o}bner bases}.
\newblock Pure and Applied Mathematics (New York). John Wiley \& Sons, Ltd.,
  Chichester, 1997.

\bibitem{Franz_Winkler:1996}
Franz Winkler.
\newblock {\em Polynomial algorithms in computer algebra}.
\newblock Texts and Monographs in Symbolic Computation. Springer-Verlag,
  Vienna, 1996.

\bibitem{Wolfram_Decker_Christoph_Lossen:2006}
Wolfram Decker and Christoph Lossen.
\newblock {\em Computing in algebraic geometry}, volume~16 of {\em Algorithms
  and Computation in Mathematics}.
\newblock Springer-Verlag, Berlin; Hindustan Book Agency, New Delhi, 2006.
\newblock A quick start using SINGULAR.

\bibitem{Wolfram_Decker_Gerhard_Pfister:2013}
Wolfram Decker and Gerhard Pfister.
\newblock {\em A first course in computational algebraic geometry}.
\newblock African Institute of Mathematics (AIMS) Library Series. Cambridge
  University Press, Cambridge, 2013.

\bibitem{Martin_Kreuzer_Lorenzo_Robbiano:2000}
Martin Kreuzer and Lorenzo Robbiano.
\newblock {\em Computational commutative algebra. 1}.
\newblock Springer-Verlag, Berlin, 2000.

\bibitem{Martin_Kreuzer_Lorenzo_Robbiano:2005}
Martin Kreuzer and Lorenzo Robbiano.
\newblock {\em Computational commutative algebra. 2}.
\newblock Springer-Verlag, Berlin, 2005.

\bibitem{Martin_Kreuzer_Lorenzo_Robbiano:2008}
Martin Kreuzer and Lorenzo Robbiano.
\newblock {\em Computational commutative algebra 1}.
\newblock Springer-Verlag, Berlin, 2008.
\newblock Corrected reprint of the 2000 original.

\bibitem{Martin_Kreuzer_Lorenzo_Robbiano:2016}
Martin Kreuzer and Lorenzo Robbiano.
\newblock {\em Computational linear and commutative algebra}.
\newblock Springer, Cham, 2016.

\bibitem{Macaulay2:Grayson:Stillman}
Daniel~R. Grayson and Michael~E. Stillman.
\newblock Macaulay2, a software system for research in algebraic geometry.
\newblock Available at \url{https://math.uiuc.edu/Macaulay2/}.

\bibitem{Bernd_Sturmfels:2002}
Bernd Sturmfels.
\newblock {\em Solving systems of polynomial equations}, volume~97 of {\em CBMS
  Regional Conference Series in Mathematics}.
\newblock Published for the Conference Board of the Mathematical Sciences,
  Washington, DC; by the American Mathematical Society, Providence, RI, 2002.

\bibitem{Invariant_Theory_Sturmfels:1993}
Bernd Sturmfels.
\newblock {\em Algorithms in invariant theory}.
\newblock Texts and Monographs in Symbolic Computation. Springer-Verlag,
  Vienna, 1993.

\bibitem{Invariant_Theory_Sturmfels:2008}
Bernd Sturmfels.
\newblock {\em Algorithms in invariant theory}.
\newblock Texts and Monographs in Symbolic Computation. Springer Wien, New
  York, Vienna, second edition, 2008.

\bibitem{Karin_Gatermann:2000}
Karin Gatermann.
\newblock {\em Computer algebra methods for equivariant dynamical systems},
  volume 1728 of {\em Lecture Notes in Mathematics}.
\newblock Springer-Verlag, Berlin, 2000.

\bibitem{Elkadi:Mourrain:2007}
Mohamed Elkadi and Bernard Mourrain.
\newblock {\em Introduction à la résolution des systèmes polynomiaux}.
\newblock Mathématiques et Applications 59, 2007.

\bibitem{AECF:2017}
Alin Bostan, Frédéric Chyzak, Marc Giusti, Romain Lebreton, Grégoire Lecerf,
  Bruno Salvy, and Éric Schost.
\newblock {\em Algorithmes Efficaces en Calcul Formel}.
\newblock HAL Archives Ouvertes, 2017.

\bibitem{Algebraic_Statistics_Pistone_Riccomagno_Wynn:2001}
Giovanni Pistone, Eva Riccomagno, and Henry~P. Wynn.
\newblock {\em Algebraic statistics}, volume~89 of {\em Monographs on
  Statistics and Applied Probability}.
\newblock Chapman \& Hall/CRC, Boca Raton, FL, 2001.
\newblock Computational commutative algebra in statistics.

\bibitem{Algebraic_Statistics_Sullivant:2018}
Seth Sullivant.
\newblock {\em Algebraic statistics}, volume 194 of {\em Graduate Studies in
  Mathematics}.
\newblock American Mathematical Society, Providence, RI, 2018.

\bibitem{Hans_J_Stetter:2004}
Hans~J. Stetter.
\newblock {\em Numerical polynomial algebra}.
\newblock Society for Industrial and Applied Mathematics (SIAM), Philadelphia,
  PA, 2004.

\bibitem{Alicia_Dickenstein_Ioannis_Z_Emiris:2005}
Alicia Dickenstein and Ioannis~Z. Emiris.
\newblock {\em Solving {P}olynomial {E}quations. {F}oundations, {A}lgorithms,
  and {A}pplications}, volume~14.
\newblock Algorithms and Computation in Mathematics, Springer, Berlin,
  Heidelberg, 2005.

\bibitem{Cox:2020:CBMS:134}
David~A. Cox.
\newblock {\em Applications of polynomial systems}, volume 134 of {\em CBMS
  Regional Conference Series in Mathematics}.
\newblock American Mathematical Society, Providence, RI, 2020.

\bibitem{CLO_UTM_2nd_ed:2005}
David~A. Cox, John Little, and Donal O'Shea.
\newblock {\em Using algebraic geometry}, volume 185 of {\em Graduate Texts in
  Mathematics}.
\newblock Springer, New York, second edition, 2005.

\bibitem{CLO_UTM_4th_ed:2015}
David~A. Cox, John Little, and Donal O'Shea.
\newblock {\em Ideals, varieties, and algorithms}.
\newblock Undergraduate Texts in Mathematics. Springer, Cham, fourth edition,
  2015.
\newblock An introduction to computational algebraic geometry and commutative
  algebra.

\bibitem{GALOIS:Cox:2012}
David~A. Cox.
\newblock {\em Galois theory}.
\newblock Pure and Applied Mathematics (Hoboken). John Wiley \& Sons, Inc.,
  Hoboken, NJ, second edition, 2012.

\bibitem{GALOIS:Stewart:2015}
Ian Stewart.
\newblock {\em Galois theory}.
\newblock CRC Press, Boca Raton, FL, fourth edition, 2015.

\bibitem{GALOIS:Tignol:2016}
Jean-Pierre Tignol.
\newblock {\em Galois' theory of algebraic equations}.
\newblock World Scientific Publishing Co. Pte. Ltd., Hackensack, NJ, second
  edition, 2016.

\bibitem{Allgower2003}
Eugene~L. Allgower and Kurt Georg.
\newblock {\em Introduction to Numerical Continuation Methods}.
\newblock SIAM, 2003.

\bibitem{Watson1990}
Layne~T. Watson.
\newblock Globally convergent homotopy algorithms for nonlinear systems of
  equations.
\newblock {\em Nonlinear Dynamics}, 1(2):143--191, Mar 1990.

\bibitem{Morgan2009}
Alexander Morgan.
\newblock {\em Solving Polynomial Systems Using Continuation for Engineering
  and Scientific Problems}.
\newblock SIAM, 2009.

\bibitem{rheinboldt1975}
Werner~C Rheinboldt.
\newblock {\em An adaptive continuation process for solving systems of
  nonlinear equations}.
\newblock University of Maryland, 1975.

\bibitem{chen2015homotopy}
Tianran Chen and Tien-Yien Li.
\newblock Homotopy continuation method for solving systems of nonlinear and
  polynomial equations.
\newblock {\em Communications in Information and Systems}, 15(2):119--307,
  2015.

\bibitem{chen2012spherical}
Tianran Chen and Tien-Yien Li.
\newblock Spherical projective path tracking for homotopy continuation methods.
\newblock {\em Communications in Information and Systems}, 12(3):195--220,
  2012.

\bibitem{10.1145/317275.317286}
Jan Verschelde.
\newblock Algorithm 795: Phcpack: A general-purpose solver for polynomial
  systems by homotopy continuation.
\newblock {\em ACM Trans. Math. Softw.}, 25(2):251–276, jun 1999.

\bibitem{10.5555/2568129}
Daniel~J. Bates, Jonathan~D. Haunstein, Andrew~J. Sommese, and Charles~W.
  Wampler.
\newblock {\em Numerically Solving Polynomial Systems with Bertini}.
\newblock Society for Industrial and Applied Mathematics, USA, 2013.

\bibitem{Numerical_Solution_2005}
Andrew Sommese and Charles Wampler.
\newblock {\em The Numerical Solution of Systems of Polynomials Arising in
  Engineering and Science}.
\newblock 03 2005.

\bibitem{Duff2018}
Timothy Duff, Cvetelina Hill, Anders Jensen, Kisun Lee, Anton Leykin, and Jeff
  Sommars.
\newblock {Solving polynomial systems via homotopy continuation and monodromy}.
\newblock {\em IMA Journal of Numerical Analysis}, 39(3):1421--1446, 04 2018.

\bibitem{moore1966interval}
Ramon~E Moore.
\newblock {\em Interval analysis}.
\newblock Prentice-Hall, 1966.

\bibitem{neumaier_1991}
Arnold Neumaier.
\newblock {\em Interval Methods for Systems of Equations}.
\newblock Encyclopedia of Mathematics and its Applications. Cambridge
  University Press, 1991.

\bibitem{10.5555/1508122}
Ramon~E. Moore, R.~Baker Kearfott, and Michael~J. Cloud.
\newblock {\em Introduction to Interval Analysis}.
\newblock Society for Industrial and Applied Mathematics, USA, 2009.

\bibitem{Kreinovich2016}
Vladik Kreinovich.
\newblock Solving equations (and systems of equations) under uncertainty: how
  different practical problems lead to different mathematical and computational
  formulations.
\newblock {\em Granular Computing}, 1(3):171--179, Sep 2016.

\bibitem{IEEE_Interval}
IEEE.
\newblock {IEEE} 1788-2015 -- {IEEE} standard for interval arithmetic.
\newblock \url{https://standards.ieee.org/standard/1788-2015.html}, 2015.

\bibitem{DBLP:reference/opt/RatschekR09}
Helmut Ratschek and Jon~G. Rokne.
\newblock Interval global optimization.
\newblock In Christodoulos~A. Floudas and Panos~M. Pardalos, editors, {\em
  Encyclopedia of Optimization, Second Edition}, pages 1739--1757. Springer,
  2009.

\bibitem{Stuber_etal}
Matthew Stuber, V.~Kumar, and P.~Barton.
\newblock Nonsmooth exclusion test for finding all solutions of nonlinear
  equations.
\newblock {\em BIT Numerical Mathematics}, 50:885--917, 12 2010.

\bibitem{Walster2007GlobalOU}
G.~William Walster and Eldon~R. Hansen.
\newblock Global optimization using interval analysis: Revised and expanded.
\newblock 2007.

\bibitem{10.2307/2952005}
Pascal~Van Hentenryck, David McAllester, and Deepak Kapur.
\newblock Solving polynomial systems using a branch and prune approach.
\newblock {\em SIAM Journal on Numerical Analysis}, 34(2):797--827, 1997.

\bibitem{Yamamura2007AnEA}
Kiyotaka Yamamura and Koki Suda.
\newblock An efficient algorithm for finding all solutions of separable systems
  of nonlinear equations.
\newblock {\em BIT Numerical Mathematics}, 47:681--691, 2007.

\bibitem{DBLP:journals/rc/Kubica13}
Bartlomiej~Jacek Kubica.
\newblock Excluding regions using sobol sequences in an interval
  branch-and-prune method for nonlinear systems.
\newblock {\em Reliab. Comput.}, 19(4):385--397, 2013.

\bibitem{intervalpatent}
G.~William Walster and Eldon~R. Hansen.
\newblock Us6915321b2: Method and apparatus for solving systems of nonlinear
  equations using interval arithmetic, 2002.

\bibitem{10.1007/978-3-540-39924-7_74}
Yves Papegay, David Daney, and Jean-Pierre Merlet.
\newblock Parallel implementation of interval analysis for equations solving.
\newblock In Jack Dongarra, Domenico Laforenza, and Salvatore Orlando, editors,
  {\em Recent Advances in Parallel Virtual Machine and Message Passing
  Interface}, pages 555--559, Berlin, Heidelberg, 2003. Springer Berlin
  Heidelberg.

\bibitem{JERRELL1998327}
Max~E. Jerrell.
\newblock Finding and verifying all solutions of a system of nonlinear
  equations.
\newblock {\em IFAC Proceedings Volumes}, 31(16):327--332, 1998.
\newblock IFAC Symposium on Computation in Economics, Finance and Engineering:
  Economic Systems, Cambridge, UK, 29 June - 1 July.

\bibitem{jerrell2002finding}
Max~E Jerrell and Wendy~A Campione.
\newblock Finding and verifying all solutions of a system of nonlinear
  equations using public domain software.
\newblock 2002.

\bibitem{eftekhari2020interval}
Tahereh Eftekhari.
\newblock Interval extensions of the halley method and its modified method for
  finding enclosures of roots of nonlinear equations.
\newblock {\em Computational Methods for Differential Equations},
  8(2):222--235, 2020.

\bibitem{didrit1998guaranteed}
Olivier Didrit, Michel Petitot, and Eric Walter.
\newblock Guaranteed solution of direct kinematic problems for general
  configurations of parallel manipulators.
\newblock {\em IEEE Transactions on Robotics and Automation}, 14(2):259--266,
  1998.

\bibitem{castellet1998algorithm}
A~Castellet and F~Thomas.
\newblock An algorithm for the solution of inverse kinematics problems based on
  an interval method.
\newblock In {\em Advances in Robot Kinematics: Analysis and Control}, pages
  393--402. Springer, 1998.

\bibitem{Yamamura1998}
Kiyotaka Yamamura, Hitomi Kawata, and Ai~Tokue.
\newblock Interval solution of nonlinear equations using linear programming.
\newblock {\em BIT Numerical Mathematics}, 38(1):186--199, Mar 1998.

\bibitem{YAMAMURA2003587}
Kiyotaka Yamamura and Tsuyoshi Fujioka.
\newblock Finding all solutions of nonlinear equations using the dual simplex
  method.
\newblock {\em Journal of Computational and Applied Mathematics},
  152(1):587--595, 2003.
\newblock Proceedings of the International Conference on Recent Advances in
  Computational Mathematics.

\bibitem{YAMAMURA2009405}
Kiyotaka Yamamura, Koki Suda, and Naoya Tamura.
\newblock Lp narrowing: A new strategy for finding all solutions of nonlinear
  equations.
\newblock {\em Applied Mathematics and Computation}, 215(1):405--413, 2009.

\bibitem{kolev1999improved}
Lubomir~V Kolev.
\newblock An improved method for global solution of non-linear systems.
\newblock {\em Reliable Computing}, 5(2):103--111, 1999.

\bibitem{Semenov_AL}
A.~Babichev, O.~Kadyrova, T.~Kashevarova, A.~Leshchenko, and A.~Semenov.
\newblock Unicalc, a novel approach to solving systems of algebraic equations.
\newblock {\em Interval Computations}, 1993, 01 1993.

\bibitem{10.1145/1132973.1132980}
Laurent Granvilliers and Fr\'{e}d\'{e}ric Benhamou.
\newblock Algorithm 852: Realpaver: An interval solver using constraint
  satisfaction techniques.
\newblock {\em ACM Trans. Math. Softw.}, 32(1):138–156, mar 2006.

\end{thebibliography}
\end{document}